\documentclass[singlespacing]{elsart}

\usepackage{graphicx}

\usepackage{amssymb}
\journal{}
\begin{document}

\begin{frontmatter}
\title{Nuclear spin-lattice relaxation from fractional wobbling in a cone}

\author{A.E. Sitnitsky},
\ead{sitnitsky@mail.knc.ru}

\address{Institute of Biochemistry and Biophysics, P.O.B. 30, Kazan
420111, Russia. e-mail: sitnitsky@mail.knc.ru, Tel. 7-843-2319037, Fax. 7-843-2927347.}

\begin{abstract}
We consider nuclear spin-lattice relaxation rate resulted from a fractional diffusion equation for anomalous rotational wobbling in a cone. The mechanism of relaxation is assumed to be due to dipole-dipole interaction of nuclear spins and is treated within the framework of the standard Bloemberger, Purcell, Pound - Solomon scheme. We consider the general case of arbitrary orientation of the cone axis relative the magnetic field. The BPP-Solomon scheme is shown to remain valid for systems with the distribution of the cone axes depending only on the tilt relative the magnetic field but otherwise being isotropic. We consider the case of random isotropic orientation of cone axes relative the magnetic field taking place in powders. Also we consider the case of their predominant orientation along or opposite the magnetic field and that of their predominant orientation transverse to the magnetic field which may be relevant for, e.g., liquid crystals. Besides we treat in details the model case of the cone axis directed along the magnetic field. The latter provides direct comparison of the limiting case of our formulas with the textbook formulas for ordinary isotropic rotational diffusion. We show that the present model enables one to obtain naturally the well known power law for Larmor frequency dependence of the spin-lattice relaxation rate. The latter is observed in some complex systems.
From this law the dependence of the fractional diffusion coefficient on the fractional index is obtained to have a rather simple functional form. The dependence of the spin-lattice relaxation rate on the cone half-width for the case of ordinary rotational diffusion yields results similar to those predicted by the model-free approach.

\end{abstract}

\begin{keyword}
NMR, spin-lattice relaxation, rotational diffusion, wobbling in a cone, model-free approach,
fractional diffusion equation.
\end{keyword}
\end{frontmatter}

\section{Introduction}
Diffusometry and relaxometry  is a traditional and well developed branch of NMR \cite{Ric80}, \cite{Kim85}, \cite{Nus88}, \cite{Kim93}, \cite{Sta94}, \cite{Sta95}, \cite{Sta951}, \cite{Fis96},
\cite{Kim97}, \cite{Kle97}, \cite{Zav98}, \cite{Zav991}, \cite{Zav99}, \cite{Kle99}, \cite{Kle02}, \cite{Kim02}, \cite{Kim04}. The so-called fractional diffusion equation (FDE) became a new theoretical tool in this field pioneered by Kimmich and coauthors \cite{Zav99} more than a decade ago.
Application of fractional calculus (for review see  \cite{Ol74}, \cite{Mil93}, \cite{Pod98}, \cite{Wes03}, \cite{Uch08} and refs. therein)
to description of fractional (or else anomalous) diffusion has been known for a long time (for review see \cite{Wes03}, \cite{Met00}, \cite{Met04}, \cite{Uch08} and refs. therein). In the present paper we deal with the particular case of the so called sub-diffusion regime.
Sub-diffusion is ubiquitous in nature  and  it originates in any fractal media due to the presence of dead ends on current ways \cite{Wes03}, \cite{Met00}, \cite{Met04}, \cite{Uch08}.
The most straightforward way to obtain sub-diffusion
mathematically is to generalize the ordinary diffusion equation
by replacing the ordinary derivative in time by the fractional
one of the order $\alpha$ ($0 < \alpha \leq 1$). The resulting FDE is a convenient
mathematical tool although a phenomenological one due to the fact
that there is no lucid and commonly accepted physical meaning of
a fractional derivative at present \cite{Wes03}, \cite{Met00}, \cite{Met04}, \cite{Uch08}. Some theoretical motivation
for introducing the FDE arises from its intrinsic relationship
with the so called continuous time random walk theory. In fact
the FDE can be derived from a generalized Langevin equation with
the memory kernel accounting for a power-law waiting time
statistics of the trapping events \cite{Met00}.

Much work is carried out at present in the field of fractional Brownian dynamics in proteins and liquids by Kneller, Hinsen and their coauthors \cite{Kne04}, \cite{Kne05}, \cite{Kne08}.
By now fractional calculus finds numerous applications in NMR diffusometry and relaxometry \cite{Zav99}, \cite{Kle99}, \cite{Kle02}, \cite{Kim02}.
In particular it is used for the description of NMR relaxation in proteins \cite{Cal08}, \cite{Cal10}.
The systematic theory of fractional calculus in NMR is developed at present mainly due to the efforts of Magin and coauthors \cite{Mag08}, \cite{Mag081}, \cite{Mag09}. However many important theoretical issues still remain open. For instance much studies were devoted to the case of translational anomalous diffusion (see \cite{Kim02} and refs. therein). In particular the generalization of the well known formula VIII.114 from \cite{Ab61} for spectral density of the correlation function for ordinary translational diffusion to the case of fractional one was obtained \cite{Sit05}, \cite{Sit08}. At the same time analogous generalization for the case of rotational diffusion has still been lacking in the literature. Although that for free rotational diffusion can be obtained trivially (basing, e.g., on the results of \cite{Cof02}, \cite{Kal04}, \cite{Ayd05}, \cite{Cof05}) it is generally of no interest for applications. In practice one is encountered with some sort of restricted rotational diffusion, e.g., wobbling in a cone treated rigorously by Wang and Pecora \cite{Wan80}. The latter model is widely used for interpreting NMR relaxation data \cite{Ric80}, \cite{Gir05}, \cite{Jar06}. The aim of the present paper is to develop the corresponding theory for fractional rotational diffusion in a cone. The approach is discussed mainly for practically important case of an isolated two-spin system comprising a hetero-nuclear pair of non-identical spins, e.g., $^{15}N - H$ in protein backbone or $^{13}C - H$ in protein side chains. However the results for homo-nuclear spin pair are also presented for the sake of completeness. From the point of view of theoretical perspectives the present paper is conceived to attain similar goals to those of \cite{Cal10}. Nevertheless the implementation is basically different. The authors of \cite{Cal10} consider the fractional Ornstein-Uhlenbeck process (sub-diffusion in a harmonic potential) and make use of the results for one-dimensional translational FDE to the case of rotational one. In contrast we consider fractional rotational diffusion in a cone within the framework of Wang and Pecora model that enables us to deal with the rotational FDE.
In our opinion both models have their merits and drawbacks.
On the one hand the harmonic potential seems to be more realistic model for restricted rotation than the cone with non-permeable boundary. On the other hand the authors of \cite{Cal10} have to assume that the internuclear vectors "fluctuate only moderately about their average direction in a molecular-fixed
frame". For wobbling in a cone we have no such restriction and we can extend the cone half-width $\theta_0$ up to the limit of isotropic rotation $\theta_0\rightarrow\pi$ to verify the coincidence of the results obtained with the known formulas.
Besides the Wang and Pecora model is exactly tractable both for ordinary diffusion and for anomalous sub-diffusion. That is why the results of rigorous treatment of the rotational FDE can serve as a touchstone for approximations by necessity invoked to in the case of more realistic potentials.

Experimental data that can be treated with the help of the theory developed in the present paper concern the magnetic field strength (or equivalently Larmor frequency) dependence of spin-lattice relaxation rate. In complex systems the latter usually exhibits a power law $\left(1/T_1\right)(\omega)=A\  \omega^{-\beta}$ where $0 < \beta \leq 1$ and $A$ is a constant. Such dependence is observed for proteins and homopolypeptides \cite{Kim85}, \cite{Nus88}, \cite{Bue99}, \cite{Bue01}, \cite{Kor01}, \cite{Kor011}, \cite{Kim04}, \cite{Kor05}, \cite{Kor06}, \cite{Kor07}, \cite{God07}, \cite{God09}, \cite{God071}, tissues \cite{Kor02}, nematic liquid crystals \cite{Leo04}, \cite{Kim04}, liquids in porous glass \cite{Sta951}, \cite{Kim04}, polymers \cite{Kim04}, etc. Purely empirical way
to obtain it is to introduce a distribution of the correlation times in the system, e.g., Cole-Davidson or Cole-Cole ones (see e.g., \cite{Jar06}, \cite{Vog08}, \cite{Sul99} and refs. therein). Usually this is done within the framework of the so-called model-free approach \cite{Lip82}, \cite{Lip821}, \cite{Bue99}, \cite{Bue01}, \cite{Mod08}, \cite{Hal09}. The generic feature of any such approach is that it does not touch upon the exponential character of the decay of the correlation function. There are more involved approaches to the problem that cast doubt on the validity of such simple basic behavior. For instance
Korb and Bryant invented an ingenious and profound method to describe the above mentioned power law of spin-lattice relaxation rate in proteins \cite{Kor01}, \cite{God07} and nematic liquid crystals \cite{Leo04}. It is based on the relaxation mechanism stipulated by spin-phonon coupling and a postulate that the protons of protein structure form a percolation network with some fractal dimensionality $d_f$. However this model is a matter of controversy and there is a competing approach to analyze experimental data \cite{Sun09}.

There is profound inherent conceptual relationship between fractals and fractional relaxation \cite{Wes03}, \cite{Met00}, \cite{Met04}, \cite{Uch08}. That is why it seems useful and interesting
to explore the problem of the power law for the spin-lattice relaxation rate in the "language" of fractional calculus. We show that the application of the single order FDE enables one to obtain the desired power law for the spin-lattice relaxation rate $\left(1/T_1\right)(\omega) \propto\omega^{-\beta}$.
Our approach leads to replacement of Debye relaxation (exponential function) by fractional relaxation (Mittag-Leffler function), i.e., to that of Lorentz functions by those
\[
 J(\omega)\propto \frac{2\tau \sin \left(\pi \alpha/2\right)(\omega  \tau)^{\alpha-1}}
 {1+ 2\cos \left(\pi \alpha/2\right)(\omega \tau)^{\alpha} +
 (\omega \tau)^{2\alpha}}
 \]
This spectral density leads in the extreme narrowing limit $\omega \tau << 1$ to the required power law for the spin-lattice relaxation rate with $\beta=1-\alpha$.
It is the well known Cole-Cole spectral density \cite{Kne04}, \cite{Kne05}, \cite{Kne08}.
The fractional index $\alpha$ phenomenologically takes into account that
the character of rotational diffusion may be different depending on the environment of the two-spin system, e.g., close packing in protein interior or loose packing at protein surface, or the influence of the environment depending the hydration extent of the protein, pressure, etc.
Our approach to rotational diffusion in application to proteins may be regarded as a complementary one to that of \cite{Kor01},
\cite{God07}. It enables to look on the problem of spin-lattice relaxation in proteins from somewhat different angle compared to the theory of Korb and Bryant.
The approach is based on a simple and visual physical model of rotational diffusion in a cone rather than on {\sl ad hoc} invented spatial distribution (fractal network) of protons in protein structure. Application of the model to different complex systems requires only the appropriate choice of the cone half-width and the value of the fractional index rather than profound modification of the structure of the theory. Our approach is in coherence with that of the paper \cite{Cal10} though as was mentioned above the implementation is basically different.

Also the advantage of fractional dynamics approach is in the fact that it enables one to reveal links with experiments and results of computer modeling beyond NMR. Prior discussing them we recall (with may be somewhat unconventional names) that in physics one usually distinguishes several types of relaxation:
1. fast relaxation (described by the exponential function),
2. slow relaxation (described by the famous stretched exponential or else  Kolraush-Williams-Watts function),
3. superslow relaxation (described by a power law function),
4. ultraslow relaxation (described by a logarithmic function), etc.
Remarkably that FDE enables one to obtain naturally any of the listed types of relaxation. The single order FDE (where the fractional index $\alpha$ is in the range $0 < \alpha \leq 1$ for the case of sub-diffusion that typically takes place in condensed matter problems)
yields the so-called Mittag-Leffler function whose intermediate regime is exactly the stretched exponential function. At asymptotically large time the stretched exponential behavior of the Mittag-Leffler function is replaced by a power law decay. The particular case $\alpha=1$ yields the normal (Debye) relaxation described by the exponential function. Finally a distributed order FDE is known to yield the ultraslow (logarithmic) relaxation at appropriate choice of the distribution $\rho(\alpha)$ over the fractional index $\alpha$.
The latter case seems to be relevant for adequate description of complex heterogeneous systems such as proteins. Indeed namely the logarithmic decay is observed in single-particle relaxation of hydrated lysozyme powder \cite{Lag09}. Nevertheless for the sake of definiteness in the present paper we restrict ourselves by the case of single order FDE and leave that of distributed order FDE for future work. Numerous interrelationships of fractional dynamics with results of molecular dynamics simulations are presented in \cite{Kne04}, \cite{Kne05}, \cite{Kne08}, \cite{Cal08}, \cite{Cal10}.

The paper is organized as follows. In Sec.2 the single order FDE for restricted rotational diffusion in a cone is used to derive the joint probability density function. In Sec.3 the latter is used to obtain the spectral densities of correlation functions for dipole-dipole interaction within the framework of the standard Bloemberger, Purcell, Pound (BPP) - Solomon scheme. In Sec.4 the dependence of the fractional diffusion coefficient on the fractional index $\alpha$ is obtained from the experimental power law for spin-lattice relaxation rate. Sec. 5-6 deal with the particular case that the cone axis is directed along the magnetic field. In Sec.5 the spin-lattice relaxation rate for hetero-nuclear spin pair is obtained. In Sec.6 that for homo-nuclear spin pair is obtained. In Sec.7 the general case of arbitrary orientation of the cone axis relative the magnetic field is considered. These results are applied to the case of isotropic random orientation (unweighted average) of cone axes relative the laboratory frame. Also a model example of predominant orientation of cone axes along or opposite the magnetic field and that of their predominant orientation transverse to the magnetic field which are considered. In Sec.8 the results are discussed and the conclusions are summarized. In Appendix A some known mathematical formulas are collected for the convenience of the reader. In Appendix B some technical details of calculations are presented.

\section{Single order fractional case for restricted rotational diffusion in a cone}
We choose a laboratory fixed frame so that its $z$ axis of the Cartesian frame  ${x,y,z}$ is directed along the constant magnetic field. The random functions $F^{(q)}$ of relative positions of two spins specified below (see (\ref{eq14})) are defined in the corresponding spherical frame ${\theta, \phi}$ given by the polar angle $\theta$ (counted from the $z$ axis) and azimuthal one $\phi$. We consider a general case that the cone axis is tilted at an angle $\psi$ relative the magnetic field. We direct the $z'$ axis of the dashed Cartesian frame ${x',y',z'}$ along the cone axis. The correlation function for wobbling in this cone we define in the corresponding spherical frame ${\theta', \phi'}$ given by the polar angle $\theta'$ (counted from the $z'$ axis) and azimuthal one $\phi'$.
Following \cite{Wan80} we consider a rod with the orientation specified by a unit
vector $\hat u$ directed along its axis with spherical polar coordinates
$\Omega' =(\theta', \phi')$. In accordance with \cite{Ab61} we following Debye assume that the rotation of the rod can be considered as that of the hard sphere with radius $a$ ($a$ is the length of the rod) in a medium with viscosity $\eta$.
For the ordinary diffusion in a cone model the rod is allowed to diffuse freely
within an empty cone with a maximum polar angle
$\theta'=\theta_0$. The symmetry axis of the
cone is taken to be the $z'$ axis. For the diffusion in a
cone model, the polar angle is restricted ($0 \leq \theta' \leq \theta_0$) but
the azimuthal angle is not ($0 \leq \phi' \leq 2\pi$).

Our aim is to consider sub-diffusion for rotational motion in a cone.
The probability density for finding the rod oriented in
$\hat u$ at time $t$, i. e., $\Psi(\hat u, t)$, obeys the single order $\alpha$ FDE for rotational motion \cite{Cof02}, \cite{Kal04}, \cite{Ayd05}, \cite{Cof05}
\begin{equation}
\label{eq1} \frac{\partial \Psi(\hat u, t)}{\partial t}=\frac{C_{\alpha}}{a^2}\ _0D_t^{1-\alpha}\Delta \Psi(\hat u, t)
\end{equation}
where $_0D_t^{1-\alpha}$ is the Riemann-Liouville fractional integral (see \cite{Ol74}, \cite{Mil93}, \cite{Pod98}, \cite{Wes03}, \cite{Wes03}, \cite{Met00}, \cite{Met04} and refs. therein), $C_{\alpha}$ is the fractional diffusion coefficient (FDC) for rotation and $\Delta$ is the angular part of the Laplace operator in polar spherical coordinates
\begin{equation}
\label{eq2} \Delta=\frac{1}{\sin^2 \theta'}\left[\sin \theta'\frac{\partial}{\partial \theta'}
\left(\sin \theta' \frac{\partial}{\partial \theta'}\right)+\frac{\partial^2}{\partial \phi'^2}\right]
\end{equation}
The FDC for rotation  $C_{\alpha}$ has the dimension $cm^2/s^{\alpha}$ and for the case $\alpha =1$ coincides with the Stokes formula
\begin{equation}
\label{eq3}  C_1 =\frac{k_B T}{8\pi a\eta}
\end{equation}
where $k_B$ is the Boltzman constant and $T$ is the temperature.

Following the combined strategy of \cite{Wan80} on the one hand and \cite{Cof02}, \cite{Kal04}, \cite{Ayd05}, \cite{Cof05} on the other hand we can write the solution of (\ref{eq1}) as follows
\[
 \Psi(\hat u, t)=
\]
\begin{equation}
\label{eq4} \sum_{n=1}^{\infty}\sum_{m=-\infty}^{\infty}
E_{\alpha}\left[-\nu_n^m(\nu_n^m+1)C_{\alpha}\mid t \mid^{\alpha}/a^2\right]
Y_{\nu_n^m}^{(m)\ \ast}\left(\Omega'(0)\right)Y_{\nu_n^m}^{(m)}\left(\Omega'(t)\right)
\end{equation}
where $E_{\alpha}(z)$ is the Mittag-Leffler function (see \cite{Ol74}, \cite{Mil93}, \cite{Pod98}, \cite{Wes03}, \cite{Wes03}, \cite{Met00}, \cite{Met04} and refs. therein), $Y_{\nu_n^m}^{(m)}\left(\Omega'\right)$ is the generalized spherical harmonics of degree $\nu_n^m$ \cite{Wan80}, the symbol $\ast$ indicates the complex conjugate and the values of $\nu_n^m$ are determined by the boundary conditions on $\theta'$ defined by our diffusion in a cone model. The boundary condition
says that there is no net change of the probability density
at the boundary of the cone, i.e.,
\begin{equation}
\label{eq5}  \frac{\partial \Psi(\hat u, t)}{\partial \theta'}
\left| {\begin{array}{l}
  \\
\theta'=\theta_0\\
 \end{array}}\right. =0
\end{equation}
The values $\nu_n^m$ are known functions of $\cos \theta_0$ \cite{Wan80},  \cite{Bau86}.
They satisfy the requirement $\nu_n^{-m}=\nu_n^m$ .
The index $n$ is defined such that $\nu_1^m < \nu_2^m < \nu_3^m <...$ for any $m$.
The detailed calculations of $\nu_n^m$ are presented in the tables \cite{Bau86}.
The values of $\nu_n^m$ for $n > 1$ increase sharply with the decrease of the confining volume.

The solution (\ref{eq4}) is subjected to the initial condition
\begin{equation}
\label{eq6}  \Psi(\hat u, 0)=\delta \left(\Omega'-\Omega'(0)\right)=
\delta\left(\cos \theta' -\cos \theta'(0)\right)\delta\left(\phi'-\phi'(0)\right)
\end{equation}
The joint probability of finding the rod with orientation
$\hat u(0)$ in solid angle $d\Omega'(0)$ at time $0$ and $\hat u(t)$ in  $d\Omega'(t)$ at
time $t$ can be written as
\[
p\left( \Omega'(t),t; \Omega'(0),0 \right)=\frac{1}{2\pi (1-\cos \theta_0)}\times
\]
\begin{equation}
\label{eq7}
\sum_{n=1}^{\infty}\sum_{m=-\infty}^{\infty}
E_{\alpha}\left[-\nu_n^m(\nu_n^m+1)C_{\alpha}\mid t \mid^{\alpha}/a^2\right]
Y_{\nu_n^m}^{(m)\ \ast}\left(\Omega'(0)\right)Y_{\nu_n^m}^{(m)}\left(\Omega'(t)\right)
\end{equation}
The latter satisfies the normalization condition
\begin{equation}
\label{eq8} \int\limits_{cone} \int\limits_{cone} p\left( \Omega'(t),t; \Omega'(0),0 \right)d\Omega'(t)d\Omega'(0)=1
\end{equation}
where the angular integrals are taken only over the volume
of the cone.

\section{NMR framework for single order fractional rotation}
At the beginning of this Sec. we recall the main facts from the general theory of spin-lattice relaxation by dipole-dipole interaction suggested by BPP and developed by Solomon \cite{Sol55}. The BPP-Solomon scheme is substantiated by more stringent Redfield's theory  (see \cite{Ab61} for detailed presentation). This scheme is developed in the frame whose $z$ axis is directed along the constant magnetic field. For the case of identical spins $I$ the contribution
to the spin-lattice relaxation rate constant due to rotational
diffusion with the spectral density at a Larmor frequency $\omega
_{L}$ of the correlation function for spherical harmonics has the
form (see VIII.76 in \cite{Ab61})
\begin{equation}
\label{eq9} \left(1/T_{1}\right)_{rotat}=\frac{3\gamma^4\hbar^2I(I+1)}{2}
\left \{ J^{(1)}(\omega _{L})+ J^{(2)}(2\omega _{L})\right \}
\end{equation}
where $\gamma$ is the gyromagnetic ratio of the nucleus, $I$ is
their spin and $\hbar$ is the Planck constant. For non-identical spins $I$ and $S$ we have four equations (see VIII.88 in \cite{Ab61})
\[
\left(1/T_{1}^{II}\right)_{rotat}=\gamma_I^2\gamma_S^2 \hbar^2 S(S+1)\times
\]
\[
\left \{\frac{1}{12}J^{(0)}\left(\omega_{L}^I-\omega_{L}^S\right)+
\frac{3}{2}J^{(1)}\left(\omega_{L}^I\right)+
\frac{3}{4}J^{(2)}\left(\omega_{L}^I+\omega_{L}^S\right)\right \}
\]
and
\[
\left(1/T_{1}^{IS}\right)_{rotat}=\gamma_I^2 \gamma_S^2 \hbar^2 I(I+1)\times
\]
\begin{equation}
\label{eq10} \left \{-\frac{1}{12}J^{(0)}\left(\omega_{L}^I-\omega_{L}^S\right)+ \frac{3}{4}J^{(2)}\left(\omega_{L}^I+\omega_{L}^S\right)\right \}
\end{equation}
Here only two equations are written out explicitly because the other two can be obtained from them by mere changing of indexes \cite{Ab61}.
To find the spectral densities $J^{(0)}(\omega)$, $J^{(1)}(\omega)$ and $J^{(2)}(\omega)$ we need to know the correlation functions $G^{(i)}(t)$ where $i=0, 1, 2\ $.

As was stressed in the previous Sec. we set $\theta'$ and $\phi'$ to be polar angles defining the direction of the axis connecting protons and $\Psi(\theta', \phi', t)=\Psi \left(\Omega', t\right)$ to be the probability of the orientation of this axis in the direction $\Omega'$ at time $t$. In the general case the axis of the cone can be tilted relative the magnetic field ($z$ axis of the laboratory fixed frame) at an arbitrary angle $\psi$. That is why at application to a realistic system the correlation function of internal motion in the cone $<F^{(i)}(0)F^{(i)}(t)>_{internal}$ has to be averaged over the orientations of cone axes with some overall distribution of the angles $f(\psi)$ characterizing the system of interest. That is the correlation function $G^{(i)}(t)$ whose spectral densities are to be substituted into (\ref{eq9}) or (\ref{eq10}) has the form
\[
G^{(i)}(t)=<<F^{(i)}(0)F^{(i)}(t)>_{internal}>_{overall}=
\]
\begin{equation}
\label{eq11}\frac{1}{2} \int \limits_{0}^{\pi} d\psi\ \sin \psi\ f(\psi )<F^{(i)}(0)F^{(i)}(t)>_{internal}
\end{equation}
To put it differently we assume that the distribution function $f(\psi, \lambda, \omega)$ (where $\psi, \lambda, \omega$ are Euler angles for rotation of the dashed Cartesian frame ${x',y',z'}$ relative the laboratory fixed one ${x,y,z}$) depends only on the Euler angle $\psi$, i.e., $f(\psi, \lambda, \omega)\equiv f(\psi)$.
It will be shown in Sec. 7 that only in this case the overall averaging provides the absence of cross-correlational functions with $q \not= q'$, i.e.,
\begin{equation}
\label{eq12} << F^{(q)}(0)F^{(q^{\prime})}(t)>_{internal}>_{overall}=\delta_{qq^{\prime}}G^{(q)}(t)
\end{equation}
where $\delta_{nm}$ is the Kronecker symbol. The latter requirement is crucial for the validity of the BPP-Solomon scheme \cite{Ab61}.
Further we consider four particular cases.

a). The cone axis is directed along the magnetic field for all cones in the system. It means that $f(\psi)=\delta (\psi)$ where $\delta (x)$ is a Dirac $\delta-$function (see Appendix B for technical details). This case is of little practical significance. However it provides direct comparison of the limiting case of our results with the textbook formulas from \cite{Ab61}. Thus it serves as a test for the validity of the present approach from the theoretical side. Besides the formulas obtained in this case without superfluous complexities are further used in more involved cases as building blocks. That is why we denote the correlation functions $G^{(i)}(t)$ for this case as basic ones $g^{(i)}(t)\equiv G^{(i)}(t)_{f(\psi)=\delta (\psi)}$. This case is considered in details in Sec.5-Sec.6.

b). For the particular case of random isotropic distribution (unweighted average) of cone axes relative the magnetic field  we have $f(\psi)=1$. This case is considered in Sec. 7.

c). As an example of the case for the cone axes to be predominantly oriented along or opposite the magnetic field we consider the model function $f(\psi)=3\cos^2 \psi$. This case is considered in Sec. 7.

d). As an example of the case for the cone axes to be predominantly oriented transverse to the magnetic field we consider the model function $f(\psi)=3/2\sin^2 \psi$. This case is considered in Sec. 7.

From now and up to the end of Sec. 6 we consider the case a)., i.e., $f(\psi)=\delta (\psi)$ where $\delta (x)$ is a Dirac $\delta-$function.
In this case each cone axis is directed along the magnetic field and we need not distinguish the dashed Cartesian frame ${x',y',z'}$ from the laboratory fixed one ${x,y,z}$. To retain the uniformity of designations for correlation function of wobbling in a cone we further use the dashed Cartesian frame and correspondingly the dashed spherical frame ${\theta', \phi'}$.
We start from the general expression for the correlation functions of arbitrary order (see VIII.13 in  \cite{Ab61}) that in our case takes the form
\[
g^{(i)}(t)\equiv G^{(i)}(t)_{f(\psi)=\delta (\psi)}= <F^{(i)}(0)F^{(i)}(t)>_{internal}=
\]
\begin{equation}
\label{eq13} \int\limits_{cone} \int\limits_{cone} F^{(i)\ \ast}\left(\Omega'(t)\right) F^{(i)}\left(\Omega'(0)\right)
p\left( \Omega'(t),t; \Omega'(0),0 \right)d\Omega'(t)d\Omega'(0)
\end{equation}
where $i=0, 1, 2$ and the angular integrals are taken only over the volume
of the cone.
We need the correlation functions $g^{(0)}(t)$, $g^{(1)}(t)$ and $g^{(2)}(t)$ in order to substitute their spectral densities in (\ref{eq9}) or (\ref{eq10}). For our case of dipole-dipole interaction of two spins separated by the distance $b$ they are defined by random functions $F^{(0)}$, $F^{(1)}$ and $F^{(2)}$ \cite{Ab61} whose relationship with associated Legendre functions $P_{2}^{(q)}\left(\cos \theta\right)$ is known (see, e.g., Appendix c. in \cite{Lan74})
\[
 F^{(0)}\left(\Omega\right)=\frac{1-3\cos^2 \theta }{b^3}=
-\frac{2}{b^3}P_{2}^{(0)}\left(\cos \theta\right)
\]
\[
 F^{(1)}\left(\Omega\right)=\frac{\sin \theta \cos \theta \exp (i\phi)}{b^3}=
\frac{1}{3b^3}P_{2}^{(1)}\left(\cos \theta\right)\exp (i\phi)
\]
\begin{equation}
\label{eq14} F^{(2)}\left(\Omega\right)=\frac{\sin^2 \theta  \exp (2i\phi)}{b^3}=
\frac{1}{3b^3}P_{2}^{(2)}\left(\cos \theta\right)\exp (i2\phi)
\end{equation}
We stress once more that in our particular case $f(\psi)=\delta (\psi)$ the $F^{(i)}\left(\Omega\right)$ in the laboratory fixed frame are identical to $F^{(i)}\left(\Omega'\right)$ in the dashed (cone-related) frame to be substituted in (\ref{eq13}).

Now we have to substitute (\ref{eq7}) and (\ref{eq14}) into (\ref{eq13}). We denote
\begin{equation}
\label{eq15} \mu=\cos \theta'
\end{equation}
so that $\mu_0=\cos \theta_0$ and introduce the associated Legendre functions $P_{\nu_n^m}^{(m)}\left(\mu\right)$
\begin{equation}
\label{eq16} Y_{\nu_n^m}^{(m)}\left(\Omega'\right)=\sqrt{\frac{1}{2\pi H_n^{(m)}}}
\exp (im\phi')P_{\nu_n^m}^{(m)}\left(\mu\right)
\end{equation}
which satisfy the orthogonality properties
\begin{equation}
\label{eq17} \int \limits_{\mu_0}^1 d\mu\  P_{\nu_{n_1}^m}^{(m)}\left(\mu\right)
P_{\nu_{n_2}^m}^{(m)}\left(\mu\right)=H_{n_1}^{(m)}\delta_{n_1, n_2}
\end{equation}
where $\delta_{n, m}$ is the Kronecker symbol ($\delta_{n, m}=1$ if $n=m$ and $\delta_{n, m}=0$ otherwise).

Making use of 1.12.1.12 and 1.12.1.9 from \cite{Pru03} (see Appendix A) respectively we obtain after straightforward calculations
\[
g^{(1)}(t)=\frac{1+\mu_0}{b^6}
\sum_{n=1}^{\infty}
E_{\alpha}\left[-\nu_n^1(\nu_n^1+1)C_{\alpha}\mid t \mid^{\alpha}/a^2\right]\frac{1}{H_n^{(1)}}\times
\]
\begin{equation}
\label{eq18} \frac{1}{(\nu_n^1+3)^2(\nu_n^1-2)^2}
\left\lbrace \left[(\nu_n^1+3)\mu_0^2-1 \right]P_{\nu_{n}^1}^{(1)}\left(\mu_0\right)-
\nu_n^1 \mu_0 P_{\nu_{n}^1+1}^{(1)}\left(\mu_0\right)\right\rbrace^2
\end{equation}
and
\[
g^{(2)}(t)=\frac{(1-\mu_0)^2(1+\mu_0)^3}{b^6}
\sum_{n=1}^{\infty}
E_{\alpha}\left[-\nu_n^2(\nu_n^2+1)C_{\alpha}\mid t \mid^{\alpha}/a^2\right]\frac{1}{H_n^{(2)}}\times
\]
\begin{equation}
\label{eq19} \frac{1}{(\nu_n^2+3)^2(\nu_n^2-2)^2}
\left\lbrace P_{\nu_{n}^2}^{(3)}\left(\mu_0\right)\right\rbrace^2
\end{equation}

The calculation of $G^{(0)}(t)$ requires the formula from \cite{Wan80}
\[
K_n^0=\int \limits_{\mu_0}^1 d\mu\  (3\mu^2-1)P_{\nu_{n}^0}^{(0)}=
4z_0 \Biggl [ \left(1-6z_0+6z_0^2\right)F\left(-\nu_{n}^0,\nu_{n}^0+1;2;z_0\right)+
\]
\begin{equation}
\label{eq20}
3z_0\left(1-2z_0\right)F\left(-\nu_{n}^0,\nu_{n}^0+1;3;z_0\right)+
2z_0^2F\left(-\nu_{n}^0,\nu_{n}^0+1;4;z_0\right)\Biggr ]
\end{equation}
where
\begin{equation}
\label{eq21} z_0=\frac{1-\mu_0}{2}
\end{equation}
and $F(a,b;c;x)$ is a hypergeometric function. Making use of (\ref{eq20}) we obtain
\begin{equation}
\label{eq22} g^{(0)}(t)=\frac{1}{(1-\mu_0)b^6}
\sum_{n=1}^{\infty}
E_{\alpha}\left[-\nu_n^0(\nu_n^0+1)C_{\alpha}\mid t \mid^{\alpha}/a^2\right]\frac{1}{H_n^{(0)}}
\left\lbrace K_n^0\right\rbrace^2
\end{equation}

We denote
\begin{equation}
\label{eq23} \tau^{(m)}_n=\left(\frac{a^2}{\nu_n^m(\nu_n^m+1)C_{\alpha}}\right)^{1/\alpha}
\end{equation}
where $m=0,1,2$.
Making use of the known Fourier transform of the Mittag-Leffler function  \cite{Kne04} we obtain
for the basic spectral densities $j^{(0)}(\omega)$, $j^{(1)}(\omega)$ and $j^{(2)}(\omega)$ of $g^{(0)}(t)$, $g^{(1)}(t)$ and $g^{(2)}(t)$ respectively
\[
j^{(0)}(\omega )=\frac{2\sin \left(\pi \alpha/2\right)}{(1-\mu_0)b^6}\times
\]
\begin{equation}
\label{eq24} \sum_{n=1}^{\infty}\frac{\tau^{(0)}_n\Bigl |\omega \tau^{(0)}_n\Bigr | ^{\alpha-1}}{1+ 2\cos \left(\pi \alpha/2\right)\Bigl |\omega \tau^{(0)}_n\Bigr |^{\alpha} +\Bigl |\omega \tau^{(0)}_n\Bigr |^{2\alpha}}\frac{1}{H_n^{(0)}}
\left\lbrace K_n^0\right\rbrace^2
\end{equation}
and
\[
j^{(1)}(\omega )=\frac{2\sin \left(\pi \alpha/2\right)}{b^6}
\sum_{n=1}^{\infty}\frac{\tau^{(1)}_n\Bigl |\omega \tau^{(1)}_n\Bigr |^{\alpha -1}}{1+ 2\cos \left(\pi \alpha/2\right)\Bigl |\omega \tau^{(1)}_n\Bigr |^{\alpha} +\Bigl |\omega \tau^{(1)}_n\Bigr |^{2\alpha}}\frac{1}{H_n^{(1)}}\times
\]
\begin{equation}
\label{eq25} \frac{(1+\mu_0)}{(\nu_n^1+3)^2(\nu_n^1-2)^2}
\left\lbrace \left[(\nu_n^1+3)\mu_0^2-1 \right]P_{\nu_{n}^1}^{(1)}\left(\mu_0\right)-
\nu_n^1 \mu_0 P_{\nu_{n}^1+1}^{(1)}\left(\mu_0\right)\right\rbrace^2
\end{equation}
and
\[
j^{(2)}(\omega )=\frac{2\sin \left(\pi \alpha/2\right)}{b^6}
\sum_{n=1}^{\infty}\frac{\tau^{(2)}_n\Bigl |\omega \tau^{(2)}_n\Bigr |^{\alpha-1}}{1+ 2\cos \left(\pi \alpha/2\right)\Bigl |\omega \tau^{(2)}_n\Bigr |^{\alpha} +\Bigl |\omega \tau^{(2)}_n\Bigr |^{2\alpha}}\times
\]
\begin{equation}
\label{eq26} \frac{(1-\mu_0)^2(1+\mu_0)^3}{H_n^{(2)}(\nu_n^2+3)^2(\nu_n^2-2)^2}
\left\lbrace P_{\nu_{n}^2}^{(3)}\left(\mu_0\right)\right\rbrace^2
\end{equation}

These spectral densities enable us to calculate any of the spin-lattice relaxation rates (\ref{eq9})-(\ref{eq10}) for the particular case that the cone axis is directed along the magnetic field. For the sake of saving room we do not write out them in the general form until the fractional diffusion coefficient $C_{\alpha}$ is specified as an explicit function of the fractional index $\alpha$. In the next Sec. we attain the latter goal on the basis of experimental data.

\section{Fractional diffusion coefficient}
As was posed in the introduction the main aim of the paper is to suggest the interpretation of the power law $\left(1/T_1\right)(\omega)=A\  \omega^{-\beta}$ (where $0 < \beta \leq 1$ and $A$ is a constant) in the Larmor frequency dependence of spin-lattice relaxation rate that is ubiquitous in complex systems.
It should be stressed that the fractional diffusion coefficient $C_{\alpha}$ is generally an unknown function of the fractional index $\alpha$. Ideally the function  $C_{\alpha}$ should be derived from some first principles for each particular system under consideration. This purpose can hardly be achieved at present level of knowledge. That is why we can pose the following inverse problem: basing on the available experimental data for the spin-lattice relaxation rate we find the unknown function $C_{\alpha}$ to satisfy the above mentioned power law. It will be shown below that for this purpose we should set
\begin{equation}
\label{eq27}  C_{\alpha}=\frac{a^2}{\tau^{\alpha}}\sin \left(\frac{\pi \alpha}{2}\right)
\end{equation}
where we denote the correlation time $\tau$ and the rotational correlation time $\tau_{rotat}$
\begin{equation}
\label{eq28} \tau =\frac{8\pi a^3\eta}{k_B T}\ \ \ \ \ \ \ \ \ \ \ \ \ \ \ \ \ \ \ \ \ \ \ \ \ \ \ \
\tau_{rotat} =\frac{4\pi a^3\eta}{3k_B T}=\frac{\tau}{6}
\end{equation}
At $\alpha=1$ we obtain (\ref{eq3}) as it must be.
Any other choice of the function $C_{\alpha}$ leads to strong dependence $\propto \sin \left[\pi (1-\beta)/2\right]$ of the constant $A$ in the power law on the parameter $\beta$ that is ruled out by the experiment.

\section{Spin-lattice relaxation rate for nuclear pair with non-identical spins}
For the particular case that the cone axis is directed along the magnetic field we have for the spectral densities to be substituted in (\ref{eq9})-(\ref{eq10}) are: $J^{(0)}(\omega )\equiv j^{(0)}(\omega )$,  $J^{(1)}(\omega )\equiv j^{(1)}(\omega )$ and $J^{(2)}(\omega )\equiv j^{(2)}(\omega )$ where $j^{(0)}(\omega )$, $j^{(1)}(\omega )$ and $j^{(2)}(\omega )$ are given by (\ref{eq24})-(\ref{eq26}).
It is worthy to note that if we identify $S$ with, e.g., $^{15}N$ from a nuclear pair of non-identical spins $^{15}N - H$
then we actually need only the formula for $\left(1/T_{1}^{II}\right)_{rotat}$ (see (\ref{eq10})) for the analysis of experimental data. That is why further we restrict ourselves only with explicit writing out the formula for this quantity. The substitution of (\ref{eq27}) into (\ref{eq24})-(\ref{eq26}) and the latter into (\ref{eq10}) yields
\[
\left( \frac{b^6}{\tau\gamma_I^2\gamma_S^2 \hbar^2 S(S+1)}\right)
\left(\frac{1}{T_{1}^{II}}\right)_{rotat}=\Bigl |\omega_{L}^I \tau \Bigr |^{\alpha-1}\sum_{n=1}^{\infty}\Biggl \{\frac{\Bigl |1-\gamma_S/\gamma_I\Bigr |^{\alpha-1}\left\lbrace K_n^0\right\rbrace^2}{6(1-\mu_0)H_n^{(0)}\nu_n^0(\nu_n^0+1)}
\times
\]
\[
\left [1+ \frac{2\cot \left(\pi \alpha/2\right)\Bigl |\left (1-\gamma_S/\gamma_I\right )\omega_{L}^I \tau \Bigr |^{\alpha}}{\nu_n^0(\nu_n^0+1)}+\frac{\Bigl |\left (1-\gamma_S/\gamma_I\right )\omega_{L}^I \tau \Bigr |^{2\alpha}}{\left[\nu_n^0(\nu_n^0+1)\sin \left(\pi \alpha/2\right)\right]^{2}}\right]^{-1}+
\]
\[
\frac{\left[\nu_n^1(\nu_n^1+1)\right]^{-1}}{1+ 2\cot \left(\pi \alpha/2\right)\Bigl |\omega_{L}^I\tau\Bigr |^{\alpha} \left[\nu_n^1(\nu_n^1+1)\right]^{-1}+\Bigl |\omega_{L}^I \tau\Bigl |^{2\alpha}\left[\nu_n^1(\nu_n^1+1)\sin \left(\pi \alpha/2\right)\right]^{-2}}\times
\]
\[
\frac{3(1+\mu_0)}{H_n^{(1)}(\nu_n^1+3)^2(\nu_n^1-2)^2}
\left\lbrace \left[(\nu_n^1+3)\mu_0^2-1 \right]P_{\nu_{n}^1}^{(1)}\left(\mu_0\right)-
\nu_n^1 \mu_0 P_{\nu_{n}^1+1}^{(1)}\left(\mu_0\right)\right\rbrace^2+
\]
\[
\left [1+ \frac{ 2\cot \left(\pi \alpha/2\right)\Bigl |\left (1+\gamma_S/\gamma_I\right )\omega_{L}^I \tau \Bigr |^{\alpha}}{\left[\nu_n^2(\nu_n^2+1)\right]} +\frac{\Bigl |\left (1+\gamma_S/\gamma_I\right )\omega_{L}^I \tau \Bigr |^{2\alpha}}{\left[\nu_n^2(\nu_n^2+1)\sin \left(\pi \alpha/2\right)\right]^{2}}\right ]^{-1}\times
\]
\begin{equation}
\label{eq29} \frac{3(1-\mu_0)^2(1+\mu_0)^3\Bigl |1+\gamma_S/\gamma_I\Bigr |^{\alpha-1}}{2H_n^{(2)}(\nu_n^2+3)^2(\nu_n^2-2)^2\nu_n^2(\nu_n^2+1)}
\left\lbrace P_{\nu_{n}^2}^{(3)}\left(\mu_0\right)\right\rbrace^2\Biggr \}
\end{equation}

This formula describes the spin-lattice relaxation rate from restricted rotational diffusion in a cone for the particular case of the cone axis to be directed along the magnetic field. The series in this formula is well convergent. That is why in practice it is sufficient to restrict oneself only by several initial terms in it.

In the extreme narrowing limit $\Bigl | \omega_{L}^I \tau \Bigr | << 1$ the formula (\ref{eq29}) yields the required power law
\begin{equation}
\label{eq30} \left(\frac{1}{T_{1}^{II}}\right)_{rotat}\approx
\frac{\widetilde A}{\Bigl |\omega_{L}^I \tau\Bigr |^{\beta}}
\end{equation}
with
\begin{equation}
\label{eq31} \beta=1-\alpha
\end{equation}
Here $\widetilde A$ is related to the experimental value $A$ by obvious relationship $A=\widetilde A/\tau^{\beta}$. The expression for $\widetilde A$ is
\[
\widetilde A=\frac{\tau \gamma_I^2\gamma_S^2 \hbar^2 S(S+1)}{b^6}\sum_{n=1}^{\infty}\Biggl \{\frac{\Bigl |1-\gamma_S/\gamma_I\Bigr |^{\alpha-1}\left\lbrace K_n^0\right\rbrace^2}{6(1-\mu_0)H_n^{(0)}\nu_n^0(\nu_n^0+1)}+
\]
\[
\frac{3(1+\mu_0)\left[\nu_n^1(\nu_n^1+1)\right]^{-1}}{H_n^{(1)}(\nu_n^1+3)^2(\nu_n^1-2)^2}
\left\lbrace \left[(\nu_n^1+3)\mu_0^2-1 \right]P_{\nu_{n}^1}^{(1)}\left(\mu_0\right)-
\nu_n^1 \mu_0 P_{\nu_{n}^1+1}^{(1)}\left(\mu_0\right)\right\rbrace^2+
\]
\begin{equation}
\label{eq32} \frac{3(1-\mu_0)^2(1+\mu_0)^3\Bigl |1+\gamma_S/\gamma_I\Bigr |^{\alpha-1}}{2H_n^{(2)}(\nu_n^2+3)^2(\nu_n^2-2)^2\nu_n^2(\nu_n^2+1)}
\left\lbrace P_{\nu_{n}^2}^{(3)}\left(\mu_0\right)\right\rbrace^2\Biggr \}
\end{equation}
Equation (\ref{eq32}) yields the relationship of the value $\widetilde A$ with the half-width of the cone $\theta_0$, i.e., $\mu_0$ (see (\ref{eq15})). However as $\nu_n^m$, $K_n^{m}$ and $H_n^{(m)}$ depend on $\mu_0$ this relationship can not be solved as an equation for the unknown $\theta_0$ from the known value of the parameter $\widetilde A$. In practice it is only feasible to seek the required value of $ \widetilde A$ with the help of varying $\theta_0$ by the trial-and-error method.

The choice of the fractional diffusion coefficient $C_{\alpha}$ in the form (\ref{eq27}) enables us to eliminate strong dependence $\propto \sin \left(\pi \alpha/2\right)$ of $\widetilde A$ on the factional index $\alpha$. However $\widetilde A$ given by (\ref{eq32}) still exhibits some weak dependence on $\alpha$. The latter is negligible.

\section{Spin-lattice relaxation rate for nuclear pair with identical spins}
For the case of identical spins the substitution of (\ref{eq27}) into (\ref{eq24})-(\ref{eq26}) and the latter into (\ref{eq9}) yields
\[
\left( \frac{b^6}{\tau\gamma^4 \hbar^2 I(I+1)}\right)
\left(\frac{1}{T_{1}}\right)_{rotat}=3\Bigl |\omega_{L} \tau \Bigr |^{\alpha-1}\sum_{n=1}^{\infty}\Biggl \{\frac{1}{\nu_n^1(\nu_n^1+1)}\times
\]
\[
\frac{1}{1+ 2\cot \left(\pi \alpha/2\right)\Bigl |\omega_{L}\tau\Bigr |^{\alpha} \left[\nu_n^1(\nu_n^1+1)\right]^{-1}+\Bigl |\omega_{L} \tau\Bigl |^{2\alpha}\left[\nu_n^1(\nu_n^1+1)\sin \left(\pi \alpha/2\right)\right]^{-2}}\times
\]
\[
\frac{(1+\mu_0)}{H_n^{(1)}(\nu_n^1+3)^2(\nu_n^1-2)^2}
\left\lbrace \left[(\nu_n^1+3)\mu_0^2-1 \right]P_{\nu_{n}^1}^{(1)}\left(\mu_0\right)-
\nu_n^1 \mu_0 P_{\nu_{n}^1+1}^{(1)}\left(\mu_0\right)\right\rbrace^2+
\]
\[
\frac{1}{1+  2\cot \left(\pi \alpha/2\right)\Bigl |2\omega_{L} \tau \Bigr |^{\alpha}\left[\nu_n^2(\nu_n^2+1)\right]^{-1} +\Bigl |2\omega_{L} \tau \Bigr |^{2\alpha}\left[\nu_n^2(\nu_n^2+1)\sin \left(\pi \alpha/2\right)\right]^{-2}}\times
\]
\begin{equation}
\label{eq33} \frac{(1-\mu_0)^2(1+\mu_0)^3 2^{\alpha-1}}{H_n^{(2)}(\nu_n^2+3)^2(\nu_n^2-2)^2\nu_n^2(\nu_n^2+1)}
\left\lbrace P_{\nu_{n}^2}^{(3)}\left(\mu_0\right)\right\rbrace^2\Biggr \}
\end{equation}
It will be shown later that in the limit of ordinary ($\alpha=1$) isotropic ($\theta_0\rightarrow \pi$) rotational diffusion this formula yields the well known result VIII.105 from \cite{Ab61}.

\section{Arbitrary orientation of the cone axis relative the magnetic field}
In the general case of arbitrary tilted cone axis relative the magnetic field we need two frames (see Sec.2 and Sec.3). The laboratory fixed frame has the $z$ axis directed along the magnetic field while the dashed (cone-related) frame has the $z'$ axis directed along the cone axis. The angle between the cone axis and the magnetic field is $\psi$. In (\ref{eq13}) we carry out internal averaging over the rotation in the cone in the dashed (cone-related) frame. That is why we need the transformation of the $F^{(i)}\left(\Omega\right)$ given by (\ref{eq14}) in the laboratory fixed frame into those in the dashed (cone-related) frame.

As is well known a rotation of one frame relative the other is most conveniently described by Euler angles. We choose the angle $\psi$ as the first Euler angle ($0 \leq \psi \leq \pi$). We denote two others Euler angles as $\lambda$ (that between the $x$-axis and the so-called $N$-line (node-line) $0 \leq \lambda \leq 2\pi$) and $\omega$ (that between the $N$-line and the $x'$-axis $0 \leq \omega \leq 2\pi$). The formula for transformation of the generalized spherical harmonics at transition from the frame $\{\phi, \theta\}$ to that $\{\phi ', \theta '\}$ obtained by rotation of the $z$ axis by the angle $\psi$ is \cite{Jef65}, \cite{Aks86}
\begin{equation}
\label{eq34} P_2^{(q)}(\cos \theta)\exp (i q \phi) =\sum_{s=-n}^{n} R_{2,q}^{(s)}(\psi) P_2^{(s)}(\cos \theta')\exp \left[i s \phi'+ i q\lambda + i s \omega\right]
\end{equation}
where
\[
R_{2,q}^{(s)}(\psi)=\sum_{r=max(0,-q-s)}^{min(2-q,2-s)} (-i)^{4-2r-q-s}\times
\]
\begin{equation}
\label{eq35} \frac{(2+q)!(2-s)!}{r!(2-q-r)!(2-s-r)!(q+s+r)!}\left(\cos \frac{\psi}{2}\right)^{q+s+2r} \left(\sin \frac{\psi}{2}\right)^{4-q-s-2r}
\end{equation}
One can see that if the distribution function $f(\psi, \lambda, \omega)$ characterizing a system of interest depends only on the angle $\psi$, i.e., $f(\psi, \lambda, \omega)\equiv f(\psi)$ then at overall averaging
\begin{equation}
\label{eq36} <...>_{overall}=\frac{1}{8\pi^2}\int \limits_{0}^{\pi} d\psi \int \limits_{0}^{2\pi} d\lambda \int \limits_{0}^{2\pi} d\omega \ \sin \psi\ f(\psi, \lambda, \omega)...
\end{equation}
we have the factors
\begin{equation}
\label{eq37} \int \limits_{0}^{2\pi} d\lambda\ \exp \left[i(q-q^{\prime})\lambda\right] =2\pi \delta_{qq^{\prime}}
\end{equation}
\begin{equation}
\label{eq38} \int \limits_{0}^{2\pi} d\omega\ \exp \left[i(s-s^{\prime})\omega\right] =2\pi \delta_{ss^{\prime}}
\end{equation}
It is namely the identity (\ref{eq37}) that provides the applicability of the formula (\ref{eq12}). The latter is crucial for the validity of the BPP-Solomon scheme \cite{Ab61}. The overall averaging takes the form
\begin{equation}
\label{eq39} <...>_{overall}=\frac{1}{2} \int \limits_{0}^{\pi} d\psi\ \sin \psi\ f(\psi )...
\end{equation}
and total (overall $+$ internal) averaging is given by (\ref{eq11}) in this case. The function $f(\psi )$ must be normalized so that
\begin{equation}
\label{eq40} \frac{1}{2} \int \limits_{0}^{\pi} d\psi\ \sin \psi\ f(\psi )=1
\end{equation}

We denote
\[
h_{(0)}^{(2)}=\int \limits_{0}^{\pi} d\psi\ \sin \psi f(\psi)\left(\cos \frac{\psi}{2}\right)^4\left(\sin \frac{\psi}{2}\right)^4
\]
\[
h_{(1)}^{(2)}=\int \limits_{0}^{\pi} d\psi\ \sin \psi f(\psi)\left(\cos \frac{\psi}{2}\right)^2\left(\sin \frac{\psi}{2}\right)^2
\left[\left(\cos \frac{\psi}{2}\right)^4+\left(\sin \frac{\psi}{2}\right)^4\right]
\]
\[
h_{(2)}^{(2)}=\int \limits_{0}^{\pi} d\psi\ \sin \psi f(\psi)
\left[\left(\cos \frac{\psi}{2}\right)^8+\left(\sin \frac{\psi}{2}\right)^8\right]
\]
\[
h_{(0)}^{(1)}=\int \limits_{0}^{\pi} d\psi\ \sin^3\psi \cos^2\psi f(\psi)
\]
\[
h_{(1)}^{(1)}=\int \limits_{0}^{\pi} d\psi\ \sin \psi f(\psi)\Biggl\lbrace\frac{1}{4}\sin^4\psi-\cos^2\psi\sin^2\psi+
\]
\[
\cos^2\psi \left[\left(\sin\frac{\psi}{2}\right)^4+\left(\cos\frac{\psi}{2}\right)^4\right]\Biggr\rbrace
\]
\[
h_{(2)}^{(1)}=\int \limits_{0}^{\pi} d\psi\ \sin^3\psi f(\psi)\left[\left(\sin\frac{\psi}{2}\right)^4+\left(\cos\frac{\psi}{2}\right)^4\right]
\]
\[
h_{(0)}^{(0)}=\int \limits_{0}^{\pi} d\psi\ \sin \psi f(\psi)\left(\cos^2\psi-\frac{1}{2}\sin^2\psi\right)^2
\]
\[
h_{(1)}^{(0)}=h_{(0)}^{(1)}
\]
\begin{equation}
\label{eq41} h_{(2)}^{(0)}=\int \limits_{0}^{\pi} d\psi\ \sin^5 \psi f(\psi)
\end{equation}
After lengthy but straightforward calculations we obtain
\begin{equation}
\label{eq42} J^{(2)}(\omega)=2h_{(0)}^{(2)}j^{(0)}(\omega)+8h_{(1)}^{(2)}j^{(1)}(\omega)+\frac{1}{2}h_{(2)}^{(2)}j^{(2)}(\omega)
\end{equation}
\begin{equation}
\label{eq43} J^{(1)}(\omega)=\frac{1}{2}\left[\frac{1}{4}h_{(0)}^{(1)}j^{(0)}(\omega)+h_{(1)}^{(1)}j^{(1)}(\omega)+\frac{1}{4}h_{(2)}^{(1)}j^{(2)}(\omega)\right]
\end{equation}
\begin{equation}
\label{eq44} J^{(0)}(\omega)=\frac{1}{2}h_{(0)}^{(0)}j^{(0)}(\omega)+9h_{(1)}^{(0)}j^{(1)}(\omega)+\frac{45}{16}h_{(2)}^{(0)}j^{(2)}(\omega)
\end{equation}
where $j^{(0)}(\omega)$, $j^{(1)}(\omega)$ and $j^{(2)}(\omega)$ are given by (\ref{eq24})-(\ref{eq26}). For the case of the fractional diffusion coefficient (\ref{eq24}) consistent with the power law for the spin-lattice relaxation rate the explicit form of the basic spectral densities $j^{(0)}(\omega)$, $j^{(1)}(\omega)$ and $j^{(2)}(\omega)$ is
\[
j^{(0)}(\omega)=\frac{2\tau\Bigl |\omega \tau \Bigr |^{\alpha-1}}{b^6(1-\mu_0)}\sum_{n=1}^{\infty}\frac{1}{H_n^{(0)}\nu_n^0(\nu_n^0+1)}\left\lbrace K_n^0\right\rbrace^2
\times
\]
\begin{equation}
\label{eq45}
\left [1+ \frac{2\cot \left(\pi \alpha/2\right)\Bigl |\omega \tau \Bigr |^{\alpha}}{\nu_n^0(\nu_n^0+1)}+\frac{\Bigl |\omega \tau \Bigr |^{2\alpha}}{\left[\nu_n^0(\nu_n^0+1)\sin \left(\pi \alpha/2\right)\right]^{2}}\right ]^{-1}
\end{equation}
\[
j^{(1)}(\omega)=\frac{2\tau\Bigl |\omega \tau \Bigr |^{\alpha-1}(1+\mu_0)}{b^6}\sum_{n=1}^{\infty}\frac{(\nu_n^1+3)^{-2}(\nu_n^1-2)^{-2}}{H_n^{(1)}\nu_n^1(\nu_n^1+1)}
\times
\]
\[
\left\lbrace \left[(\nu_n^1+3)\mu_0^2-1 \right]P_{\nu_{n}^1}^{(1)}\left(\mu_0\right)-
\nu_n^1 \mu_0 P_{\nu_{n}^1+1}^{(1)}\left(\mu_0\right)\right\rbrace^2
\times
\]
\begin{equation}
\label{eq46}
\left [1+ \frac{2\cot \left(\pi \alpha/2\right)\Bigl |\omega \tau \Bigr |^{\alpha}}{\nu_n^1(\nu_n^1+1)}+\frac{\Bigl |\omega \tau \Bigr |^{2\alpha}}{\left[\nu_n^1(\nu_n^1+1)\sin \left(\pi \alpha/2\right)\right]^{2}}\right ]^{-1}
\end{equation}
\[
j^{(2)}(\omega)=\frac{2\tau\Bigl |\omega \tau \Bigr |^{\alpha-1}(1-\mu_0)^2(1+\mu_0)^3}{b^6}\sum_{n=1}^{\infty}\frac{(\nu_n^2+3)^{-2}(\nu_n^2-2)^{-2}}{H_n^{(2)}\nu_n^2(\nu_n^2+1)}
\times
\]
\begin{equation}
\label{eq47}
\left\lbrace P_{\nu_{n}^2}^{(3)}\left(\mu_0\right)\right\rbrace^2\left [1+ \frac{2\cot \left(\pi \alpha/2\right)\Bigl |\omega \tau \Bigr |^{\alpha}}{\nu_n^2(\nu_n^2+1)}+\frac{\Bigl |\omega \tau \Bigr |^{2\alpha}}{\left[\nu_n^2(\nu_n^2+1)\sin \left(\pi \alpha/2\right)\right]^{2}}\right ]^{-1}
\end{equation}

For practical application of the theory we consider three cases.

1. For the particular case of random isotropic distribution (unweighted average) of cone axes relative the magnetic field we have $f(\psi)=1$. In this case we obtain: $h_{(0)}^{(2)}=1/15$; $h_{(1)}^{(2)}=1/5$; $h_{(2)}^{(2)}=4/5$; $h_{(0)}^{(1)}=4/15$; $h_{(1)}^{(1)}=8/15$; $h_{(2)}^{(1)}=4/5$; $h_{(0)}^{(0)}=2/5$; $h_{(1)}^{(0)}=4/15$; $h_{(2)}^{(0)}=16/15$.

2. As an example of the case for the cone axes to be predominantly oriented along or opposite the magnetic field we consider the model function $f(\psi)=3\cos^2 \psi$. In this case we obtain: $h_{(0)}^{(2)}=1/35$; $h_{(1)}^{(2)}=1/7$; $h_{(2)}^{(2)}=44/35$; $h_{(0)}^{(1)}=12/35$; $h_{(1)}^{(1)}=28/35$; $h_{(2)}^{(1)}=20/35$; $h_{(0)}^{(0)}=22/35$; $h_{(1)}^{(0)}=12/35$; $h_{(2)}^{(0)}=16/35$.\\

3. As an example of the case for the cone axes to be predominantly oriented transverse to the magnetic field we consider the model function $f(\psi)=3/2\sin^2 \psi$. In this case we obtain: $h_{(0)}^{(2)}=3/35$; $h_{(1)}^{(2)}=8/35$; $h_{(2)}^{(2)}=20/35$; $h_{(0)}^{(1)}=8/35$; $h_{(1)}^{(1)}=14/35$; $h_{(2)}^{(1)}=32/35$; $h_{(0)}^{(0)}=10/35$; $h_{(1)}^{(0)}=8/35$; $h_{(2)}^{(0)}=48/35$.\\

Making use of (\ref{eq45})-(\ref{eq47}) and explicit values of $h_{(i)}^{(q)}$ for these two cases enables us to calculate the spectral densities (\ref{eq42})-(\ref{eq44}) to be inserted in BPP-Solomon formulas (\ref{eq9})-(\ref{eq10}). We do not write out explicitly the corresponding expressions for the spin-lattice relaxation rate to save room. However in the next Sec. we present the corresponding figures for spin-lattice relaxation rate for nuclear pair with non-identical spins for both cases.

\section{Results and discussion}
For $^{15}N - H$ nuclear pair of non-identical spins, we identify $S$ with $^{15}N$ nucleus and $I$ with $H$ one. Thus $\gamma_S=-2712\  rad\ s^{-1}\ Gauss^{-1}$ and $\gamma_I=26753\ rad\ s^{-1}\ Gauss^{-1}$ so that $\gamma_S/\gamma_I=-0.101372$. For
$^{13}C - H$ nuclear pair $\gamma_S=6728\  rad\ s^{-1}\ Gauss^{-1}$ so that $\gamma_S/\gamma_I=0.251486$.
The righthand side in the formula (\ref{eq29}) depends on the parameters characterizing the nuclear pair (namely on the gyromagnetic ratios $\gamma_S$ and $\gamma_I$ of our pair of non-identical spins). To plot the spin-lattice relaxation rate with the help of (\ref{eq29}) one has to choose the particular nuclear pair explicitly. That is why to be specific we choose the $^{15}N - H$ nuclear pair of non-identical spins.
In Fig. 1 the dependence of the spin-lattice relaxation rate on Larmor frequency in the cone with $\theta_0=55^\circ$ (chosen as an illustrative example) for the case of cone axes directed along the magnetic field at different values of the fractional index $\alpha$ is depicted. In Fig. 2 the dependence of the spin-lattice relaxation rate on fractional index $\alpha$ in the cone with $\theta_0=55^\circ$ for different values of Larmor frequency is depicted.
In Fig. 3 the same for $\theta_0=5^\circ$ is depicted as an example of the case of a narrow cone.
In Fig. 4 the dependence of the spin-lattice relaxation rate on the cone half-width $\theta_0$ for the case of cone axes directed along the magnetic field is depicted at $\omega_L \tau = 0.1$ for different values of the fractional index $\alpha$. In Fig. 5 the dependence of the spin-lattice relaxation rate on the cone half-width $\theta_0$ for the case of cone axes directed along the magnetic field is depicted at $\alpha=0.2$ ($\beta=0.8$ that is a typical experimental value) for different values of Larmor frequency.

In Fig. 6 the spin-lattice relaxation rate for identical spins obtained with the help of (\ref{eq33}) for the case of cone axes directed along the magnetic field is depicted as a function of the cone half-width $\theta_0$ at $\omega \tau = 0.1$ for different values of the fractional index $\alpha$. From this Fig. one can see that for the case of ordinary ($\alpha=1$) isotropic ($\theta_0=\pi$) rotational diffusion in the limit of extreme narrowing $\omega_{L} \tau << 1 $ the corresponding curve tends to the value $0.33$. Thus the formula (\ref{eq33}) yields
\begin{equation}
\label{eq48} \lim_{\theta_0\to\pi}\left( \frac{b^6}{\tau\gamma^4 \hbar^2 I(I+1)}\right)
\left(\frac{1}{T_{1}}\right)_{rotat}\left| {\begin{array}{l}
  \\
\alpha=1\\
 \end{array}}\right. =\frac{1}{3}
\end{equation}
that taking into account (\ref{eq28}) coincides for the case with the well known formula VIII.106 from \cite{Ab61}
\[
\left(\frac{1}{T_{1}}\right)_{rotat}=\frac{2\gamma^4\hbar^2 }{b^6}I(I+1)\frac{4\pi\eta a^3}{3k_BT}
\]
In Fig. 7 the dependence of the spin-lattice relaxation rate on the cone half-width $\theta_0$ for the case of random isotropic distribution (unweighted average $f(\psi)=1$) of cone axes relative the magnetic field is depicted at $\omega_L \tau = 0.1$ for different values of the fractional index $\alpha$. In Fig.8 the dependence of the spin-lattice relaxation rate on the cone half-width $\theta_0$ for the case of the cone axes to be predominantly oriented along or opposite the magnetic field ($f(\psi)=3\cos^2 \psi$) is depicted at $\omega_L \tau = 0.1$ for different values of the fractional index $\alpha$. In Fig.9 the dependence of the spin-lattice relaxation rate on the cone half-width $\theta_0$ for the case of the cone axes to be predominantly oriented transverse to the magnetic field ($f(\psi)=3/2\sin^2 \psi$) is depicted at $\omega_L \tau = 0.1$ for different values of the fractional index $\alpha$.

As is shown in Sec.3 in the extreme narrowing limit $\omega \tau << 1$ the formula (\ref{eq29})
yields the required behavior $(1/T_{1})_{rotat} (\omega)\approx A\  \omega^{-\beta}$ with $\beta=1-\alpha$. Such behavior is observed experimentally in a number of systems \cite{Bue99}, \cite{Bue01}, \cite{Kor01}, \cite{God07}, \cite{Leo04}, \cite{Kim04} (see Introduction).
The typical dependence of the spin-lattice relaxation rate on the Larmor frequency is demonstrated by Fig.1 for the example of the cone with the half-width $\theta_0=55^\circ$.
By varying the half-width of the cone $\theta_0$ and the fractional index $\alpha$ we obtain rich behavior to fit any experimentally observed power law $(1/T_{1})_{rotat} (\omega)\approx A\  \omega^{-\beta}$.
In the extreme narrowing limit we get $(1/T_{1})_{rotat}\propto \left(\eta (T)/T \right)^{\alpha}$.
In the high frequency limit $\omega \tau >> 1$ we obtain from (\ref{eq29}) $(1/T_{1})_{rotat} \propto \omega^{-(1+\alpha)}$. In this range we get $(1/T_{1})_{rotat}\propto \left(T/\eta (T)\right)^{\alpha}$. The formula (\ref{eq29}) describes the intermediate behavior between these limiting cases.

As is well known in liquids $T_1$ ordinarily decreases with increasing viscosity, in some cases reaching a minimum value after which it increases with further increase in viscosity  \cite{Ab61}.
The variation of viscosity is caused by temperature $T$. For ordinary isotropic rotation correlation times are functions of temperature $\tau_n =\tau \left(\eta (T)\right)[n(n+1)]^{-1}$ (see VIII.97 in \cite{Ab61}) and the spin-lattice relaxation rate $(1/T_{1})_{rotat} \left( \tau (T)\right)$ can have a maximum as a result. In the present model we have more options. Substitution of (\ref{eq27}) into (\ref{eq23}) yields
\begin{equation}
\label{eq49} \tau^{(m)}_n=\tau \left(\eta (T)\right)\left[\nu_n^m(\nu_n^m+1)\sin \left(\pi \alpha/2\right)\right]^{-1/\alpha}
\end{equation}
where $m=0,1,2$. Thus correlation times are functions of temperature, of the cone half-width $\theta_0$ (via the values $\nu_n^m$) and of the fractional index $\alpha$. The results obtained testify that variation of each of these parameters can produce a maximum in the corresponding dependence of the spin-lattice relaxation rate. For the limit case of ordinary ($\alpha=1$) isotropic ($\theta_0 = \pi$) rotational diffusion at extreme narrowing condition the results obtained coincide with the well known textbook formula VIII.106 from \cite{Ab61}.

From Fig.2 and Fig.3 we see that spin-lattice relaxation rate exhibits a maximum in the dependence of spin-lattice relaxation rate on the fractional index $\alpha$. For a narrow cone $\theta_0 < 30^\circ$ the position of the maximum is at values of the fractional index $\alpha < 0.1$ (see Fig.3). As the cone becomes wider the position of the maximum is shifted to higher values of the fractional index $\alpha$ (see Fig.2). For a wide cone the maximum takes place at $\alpha \approx 0.2$. Our result about the increase of the spin-lattice relaxation rate with the decrease of the fractional index $\alpha$ is in accordance with the conclusion of the authors of \cite{Kor01}, \cite{God07} that the effectively reduced dimensionality of the system (in their case it is chain structure of the protein compared with $3D$ crystalline structure) leads to more efficient spin-coupling relaxation.

In Fig.4 the dependence of the spin-lattice relaxation rate on the cone half-width $\theta_0$ is depicted. It should be mentioned that (\ref{eq29}) has uncertainties at:
$\theta_0=\pi/4$ because $\nu_1^1=2.0000$; $\theta_0=\pi/2$  because $\nu_1^2=2.0000$; $\theta_0=3\pi/4$ because $\nu_2^1=2.0000$; $\theta_0=175^\circ$ because $\nu_1^2=2.0000$.
However these uncertainties are isolated, and can be safely ignored. The dependence exhibits a maximum at $\theta_0 \approx 85 \div 110^\circ$. We have a maximum at $\theta_0 \approx 85^\circ$ even for the case $\alpha=1$, i.e., for ordinary diffusion in a cone.
These maximums result from thorough investigation of the Wang-Pecora model \cite{Wan80} based on the Bauer's tables \cite{Bau86}. The maximum for ordinary diffusion in a cone at $\theta_0 \leq \pi/2$ is similar to that given by the model-free approach.

Indeed let us consider the most typical for practice case of the model-free approach when the overall motion of a macromolecule is considerably slower than the internal motion. In this case the expression of the model-free approach for the relationship of the spin-lattice relaxation rate with the order parameter $S^2$ and effective correlation time $\tau_e$ is given by  equation (37) from \cite{Lip82}
\begin{equation}
\label{eq50} \frac{1}{T_1}=aS^2+b\tau_e\left(1-S^2\right)
\end{equation}
where $a$ and $b$ are constants independent on spatial configuration accessible for internuclear vector. For ordinary wobbling in a cone the relationship between the order parameter of the model-free approach and the cone half-width $\theta_0$  is given by equation (A3) from \cite{Lip82}
\begin{equation}
\label{eq51} S=\frac{1}{2}\cos \theta_0 (1+\cos \theta_0)
\end{equation}
while that for the effective correlation time is given by equation (A4) from \cite{Lip82}
\[
\tau_e=\frac{1}{D_w\left(1-S^2\right)}\Biggl\{\cos^2\theta_0\left(1+\cos \theta_0\right)^2
\bigl\{\ln\left[(1+\cos \theta_0)/2\right]+
\]
\[
(1-\cos \theta_0)/2\bigr\}/[2(\cos \theta_0-1)]+(1-\cos \theta_0)(6+8\cos \theta_0-
\]
\begin{equation}
\label{eq52} \cos^2 \theta_0-12\cos^3 \theta_0-7\cos^4 \theta_0)/24\Biggr\}
\end{equation}
where $D_w$ is the diffusion coefficient. We denote
\begin{equation}
\label{eq53} c=\frac{b}{aD_w}
\end{equation}
Then we obtain the dependence of the spin-lattice relaxation rate on the cone half-width $\theta_0$
\[
\frac{1}{a}\frac{1}{T_1}=\frac{1}{4}\cos^2 \theta_0 (1+\cos \theta_0)^2+c\Biggl\{\cos^2\theta_0\left(1+\cos \theta_0\right)^2
\bigl\{\ln\left[(1+\cos \theta_0)/2\right]+
\]
\[
(1-\cos \theta_0)/2\bigr\}/[2(\cos \theta_0-1)]+(1-\cos \theta_0)(6+8\cos \theta_0-
\]
\begin{equation}
\label{eq54} \cos^2 \theta_0-12\cos^3 \theta_0-7\cos^4 \theta_0)/24\Biggr\}
\end{equation}
The dependence of this spin-lattice relaxation rate on the cone half-width $\theta_0$ is depicted in Fig.11 at several values of the parameter $c$. One can see close qualitative similarity of the curves with our result for the case of ordinary wobbling in a cone.

The BPP-Solomon scheme requires essential extent the isotropy (but not total isotropy) for its validity. In Sec.7 we show that in application to rotational diffusion in a cone it remains valid for systems with a distribution of cone axes depending only on the tilt  relative the magnetic field but otherwise being isotropic. This residual isotropy provides the requirement (\ref{eq37}) that is
necessary for the absence of cross-correlational functions (\ref{eq12}). The latter in turn is crucial for the validity of of the BPP-Solomon scheme. We consider the aforesaid a so important issue that would like to reiterate it in other words with complete definiteness. We develop the theory for the general case of arbitrary orientation of the cone axis relative the magnetic field (laboratory fixed frame). We show that when the cone axis is tilted at an arbitrary angle $\psi$ to the magnetic field but otherwise is oriented isotropically then at overall averaging the crucial requirement for the validity of the Bloemberger, Purcell, Pound - Solomon scheme (\ref{eq12}) (that of the absence of cross-correlational functions with $q \not= q'$) is retained. Thus we explicitly prove the consistency of combination of the textbook formulas for the BPP-Solomon scheme with rotational motion in a cone.

Practical applications of our results depend on the choice of a distribution function for overall averaging over all orientations of the cone axis with respect to the laboratory fixed frame (over the angle $\psi$). This distribution function $f(\psi)$ is a characteristic of the system of interest. Under the assumption of isotropic random orientation of cone axes relative the laboratory fixed frame the results obtained can be applied to powders. We consider this case of unweighted average over all orientations of the cone axis with respect to the laboratory fixed frame ($f(\psi)=1$) in Sec.7 and provide corresponding data for the spin-lattice relaxation rate in Fig.7. Also in Sec.7 we consider a model example of predominant orientation of the cone axis along or opposite the magnetic field ($f(\psi)=3\cos^2 \psi$) and that of their predominant orientation transverse to the magnetic field  ($f(\psi)=3/2\sin^2 \psi$). The results for these cases may be relevant for, e.g., liquid crystals. For both of these cases we also provide corresponding data for the spin-lattice relaxation rate in Fig.8. and Fig. 9 respectively. The particular case of the cone axis directed along the magnetic field ($f(\psi)=\delta (\psi)$ where $\delta (x)$ is a Dirac delta-function) is of little practical significance. However this case provides direct comparison of the limiting case of our formulas with the textbook formulas from \cite{Ab61}. Thus it serves as a test for the validity of our approach from the theoretical side. In the limit of isotropic motion (i.e., the cone half-width = $\pi$) our formulas yield those of \cite{Ab61} as it must be.

Fig.7, Fig.8 and Fig.9 show that the results for practically relevant cases of specific distributions of cone axes relative the magnetic field are qualitatively similar to those for the model case of cone axes directed along the magnetic field (Fig.4). However quantitatively the values of spin-lattice relaxation rates differ from those for the model case. For very wide cones ($\theta_0 > 100^\circ$) the results for three cases are rather close to each other because when the cone half-width is large its tilt relative the magnetic field becomes of minor importance. Appreciable difference between three cases takes place in the region of moderate cone half-widths $25^\circ < \theta_0 < 90^\circ$. The comparison of  Fig.4, Fig.7, Fig.8 and Fig.9 shows that
\[
 \left(\frac{1}{T_1}\right)_{f(\psi)=\delta (\psi)} < \left(\frac{1}{T_1}\right)_{f(\psi)=3\cos^2 \psi} < \left(\frac{1}{T_1}\right)_{f(\psi)=1} < \left(\frac{1}{T_1}\right)_{f(\psi)=3/2\sin^2 \psi}
\]
Thus the more is the contribution of the orientations of the cone axes in the system transverse to the magnetic field the more efficiently spin-lattice relaxation proceeds. To prove it explicitly we plot in Fig.10 the the dependence of the spin-lattice relaxation rate on the tilt angle of the cone axis relative the magnetic field for the cone with the half-width $\theta_0 = 55^\circ$ as an example. This case corresponds to the distribution function $f(\psi)=\delta (\psi-\psi_0)\left[\cos \left(\psi_0/2\right)\right]^{-1}$ (see Appendix B for technical details). One can see that when the cone axis is transverse the magnetic field the spin-lattice relaxation proceeds most efficiently.

In closing we touch upon the most difficult and troublesome problem of
the physical interpretation of the results obtained in the present paper. Regretfully the latter is hampered by the general flaw of the fractional calculus and the FDE that there is no lucid and commonly accepted physical meaning of
a fractional derivative at present. As a consequence there is no clear physical interpretation of the fractional index $\alpha$ and the results based on its variation. Despite the aforesaid the application of the FDE to the description of relaxation and diffusion in complex systems is certainly a mainstream among modern approaches to this problem (see, e.g., monographs \cite{Wes03}, \cite{Uch08} and reviews \cite{Met00}, \cite{Met04} and refs. therein). Slowly this approach is finding its way into NMR \cite{Zav99}, \cite{Mag08}, \cite{Mag081}, \cite{Mag09}, \cite{Sit05}, \cite{Sit08},  \cite{Cal08}, \cite{Cal10}. Notably the paper \cite{Cal10} is aimed to apply this approach to the description of rotational motion. The present paper is also aimed to attain it. In fact the FDE is a phenomenological tool that can find at least three different physical interpretations.

a. The time-dependent part of its solution is a Mittag-Leffler function that is a natural generalization of an exponent whose spectral density is the Cole-Cole one  \cite{Cal08}, \cite{Cal10}. Thus the  Mittag-Leffler function is merely a concise and elegant way to write out the integral over the exponents with the Cole-Cole distribution. The physical meaning of the fractional index in the FDE appears to be that of an empirical parameter of the Cole-Cole distribution that a practitioner in NMR or dielectric relaxometry chooses to fit experimental data.
The authors of \cite{Cal10} incline to choose this interpretation. However there are other options that can be used even when there is no distribution of relaxation times but the rod of an isolated spin pair moves on a fractal or on a lattice whose knots are traps with some distribution of waiting times.

b. The FDE is well suited to describe diffusion on a fractal \cite{Met00}. The fractional index in it is directly related to the fractal dimensionality of the system of interest.

c. The FDE can be derived from the continuous time random walk theory of the diffusion on the lattice with traps \cite{Met00}. The physical meaning of the fractional index is that of a parameter in the power-law distribution of the trapping events.

At such abundance of possibilities to choose one of them at the level of general theoretical consideration (that the present paper deals with) and to insist on it seems to be somewhat narrow-mindedly. Of course at application to particular systems one can find convincing reasons to choose a specific interpretation. In our opinion the most honest attitude at the level of the present paper is to admit that there is no commonly accepted physical meaning of the fractional index and the results based on its variation (or more exactly there are too many of them). The latter statement does not mean that the FDE is inapplicable. It should merely be perceived properly as a phenomenological tool. This is the point of view accepted in the present paper.

We refrain ourselves from further discussing this issue. The derivation of (\ref{eq1}) is strictly speaking beyond the scope of the present paper. We feel that further attempts to motivate (\ref{eq1}) from our side would be mere reiteration of the arguments presented in \cite{Kne04}, \cite{Kne05}, \cite{Kne08}, \cite{Cal08}, \cite{Cal10} within a more general context.
In any case the FDE is so well suited for phenomenological description of relaxation processes obeying fractional power laws \cite{Wes03}, \cite{Met00}, \cite{Met04}, \cite{Uch08} that it seems reasonable and timely to apply it to nuclear spin-lattice relaxation from restricted rotational diffusion and to develop a formal theory. The paper \cite{Cal10} is a step in this direction.
The present study is also a development along this line and in our opinion for the particular case of fractional wobbling in a cone the aim has been attained. It should be stressed that the absence of the power-law is merely a particular case of the present model (the value $\alpha=1$ for a parameter $\alpha$ that can take values from the range $0 < \alpha \leq 1$). Thus the absence of the power-law in an experiment does not invalidate the model. In our opinion it is useful to have a formalism that can describe well established results as a particular case and be easily generalized (by simple varying a single parameter rather than by introducing a distribution of correlation times via some integral) to describe more complex behavior at will.

We conclude that fractional wobbling in a cone is a relatively simple, exactly tractable (within the range of validity of BPP-Solomon scheme) and convenient model for phenomenological description of experimental systems exhibiting power law in Larmor frequency dependence of nuclear spin-lattice relaxation rate.

Acknowledgements.  The author is grateful to Dr. Yu.F. Zuev and R.H. Kurbanov for
helpful discussions. The work was supported by the grant from RFBR and
the programme "Molecular and Cellular Biology" of RAS.

\section{Appendix A}
Here we present two known mathematical formulas 1.12.1.12 and 1.12.1.9 for the the associated Legendre function $P_{\nu}^{\mu}\left(x\right)$ from the table of integrals \cite{Pru03}. The formula 1.12.1.12 is
\[
 \int  dx\  x(1-x^2)^{\pm \mu/2}P_{\nu}^{\mu}\left(x\right)=\frac{(1-x^2)^{\pm \mu/2}}{(\nu \pm \mu +2)(\nu \mp \mu -1)}\times
\]
\[
\left \{ \left[(\nu \pm \mu +2)x^2-1\right]P_{\nu}^{\mu}\left(x\right)+(\mu - \nu -1)xP_{\nu+1}^{\mu}\left(x\right)\right\}
\]
The formula 1.12.1.9 is
\[
 \int  dx\  (1-x^2)^{\mu/2}P_{\nu}^{\mu}\left(x\right)=\frac{(1-x^2)^{(\mu+1)/2}}{(\nu - \mu)(\nu +\mu +1)}P_{\nu}^{\mu+1}\left(x\right)
\]

\section{Appendix B}
The polar angle $\psi$ at operations with the distribution function $f(\psi)$ imposes some peculiarities in treating the case of Dirac $\delta-$function
\[
f(\psi)=\frac{1}{\cos \left(\psi_0/2\right)}\delta (\psi-\psi_0)
\]
We stress that the case of the cone angle oriented along the magnetic field $f(\psi)=\delta (\psi)$ considered in Sec. 4- Sec.6 is a particular case of this distribution function corresponding to the value $\psi_0=0$. First let us prove the normalization requirement (\ref{eq40}). We have (taking into account that $\delta (2z)=\delta (z)/2$)
\[
\frac{1}{2} \int \limits_{0}^{\pi} d\psi\ \sin \psi\ \delta (\psi-\psi_0)=2\int \limits_{0}^{\pi} d\left(\frac{\psi}{2}\right) \sin \frac{\psi}{2} \cos \frac{\psi}{2} \delta \left(2\frac{\psi-\psi_0}{2}\right)=
\]
\[
\int \limits_{0}^{\pi/2} dx\ \sin x \cos x\ \delta \left(x-\frac{\psi_0}{2}\right)=-\int \limits_{0}^{\pi/2} d(\cos x)\ \cos x\ \delta \left(x-\frac{\psi_0}{2}\right)=
\]
\[
\int \limits_{0}^{1} dy\ y\ \delta \left(\arccos y - \frac{\psi_0}{2}\right)=\int \limits_{0}^{1} dy\ y\ \delta \left(y - \cos \frac{\psi_0}{2}\right)= \cos \frac{\psi_0}{2}
\]
This calculation serves as a model for operations at calculating the values of $h_{(m)}^{(n)}$ in (\ref{eq41}) with the distribution function $f(\psi)=\delta (\psi-\psi_0)\left[\cos \left(\psi_0/2\right)\right]^{-1}$. The general rule takes the form
\[
\frac{1}{\cos \left(\psi_0/2\right)}\int \limits_{0}^{\pi} d\psi\ \sin \psi\ \delta (\psi-\psi_0)q(\psi) =2q(\psi_0)
\]
where $q(\psi)$ is an arbitrary function of $\psi$.

At $\psi_0=0$ we obtain $h_{(2)}^{(2)}=2$, $h_{(1)}^{(1)}=2$ and $h_{(0)}^{(0)}=2$ while $h_{(m)}^{(n)}=0$ at $m\not=n$. Substitution of these values into (\ref{eq42})-(\ref{eq44}) yields $J^{(2)}(\omega )=j^{(2)}(\omega )$,  $J^{(1)}(\omega )=j^{(1)}(\omega )$ and  $J^{(0)}(\omega )=j^{(0)}(\omega )$ as it must be.

\newpage

\clearpage
\begin{figure}
\begin{center}
\includegraphics* [width=\textwidth] {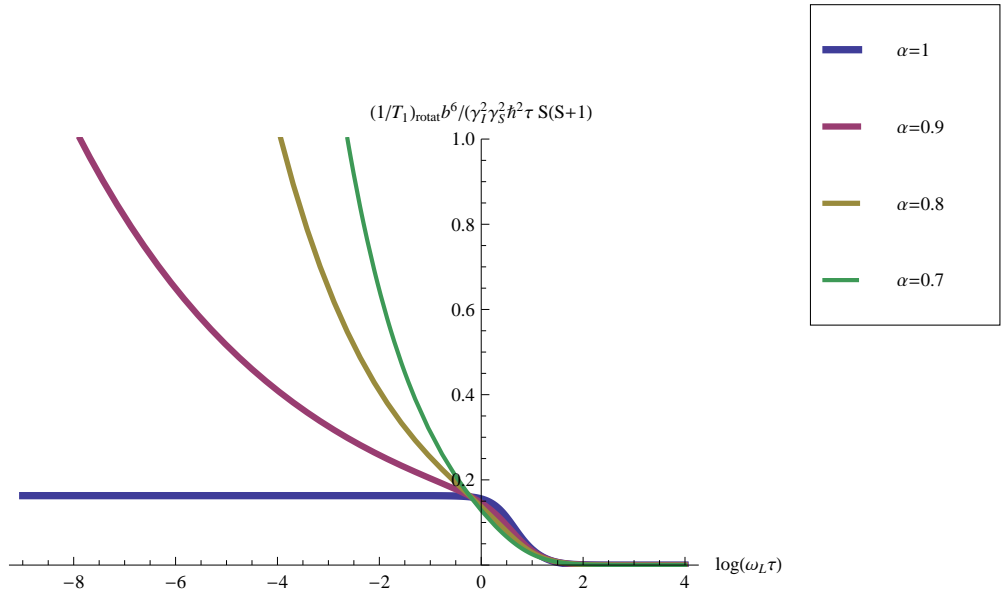}
\end{center}
\caption{Spin-lattice relaxation rate for $^{15}N - H$ nuclear pair of non-identical spins from fractional rotational diffusion for the case of cone axes directed along the magnetic field (eq. (\ref{eq29})) in the cone with $\theta_0=55^\circ$ as the function of Larmor frequency $\omega_L$
at different values of the fractional index $\alpha$: $\alpha =1$ (thick line); $\alpha =0.9$; $\alpha =0.8$; $\alpha =0.7$ (thin line).}
\label{Fig.1}
\end{figure}

\clearpage
\begin{figure}
\begin{center}
\includegraphics* [width=\textwidth] {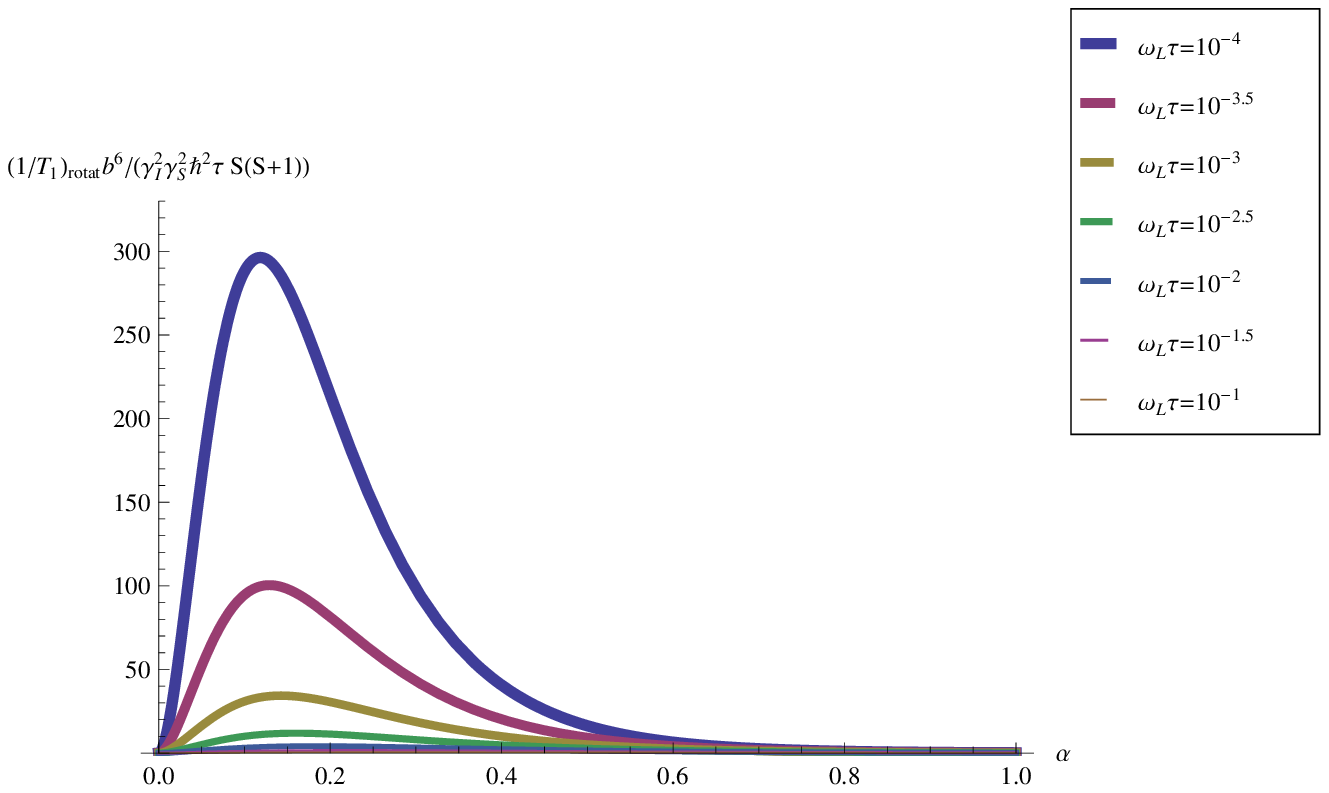}
\end{center}
\caption{Spin-lattice relaxation rate for $^{15}N - H$ nuclear pair of non-identical spins from fractional rotational diffusion for the case of cone axes directed along the magnetic field (eq. (\ref{eq29})) in the cone with $\theta_0=55^\circ$ as the function of the fractional index $\alpha$
at different values of Larmor frequency $\omega_L$: $\omega_L\tau=10^{-4}$ (thick line); $\omega_L\tau=10^{-3.5}$; $\omega_L\tau=10^{-3}$; $\omega_L\tau=10^{-2.5}$; $\omega_L\tau=10^{-2}$; $\omega_L\tau=10^{-1.5}$; $\omega_L\tau=10^{-1}$ (thin line).}
\label{Fig.2}
\end{figure}

\clearpage
\begin{figure}
\begin{center}
\includegraphics* [width=\textwidth] {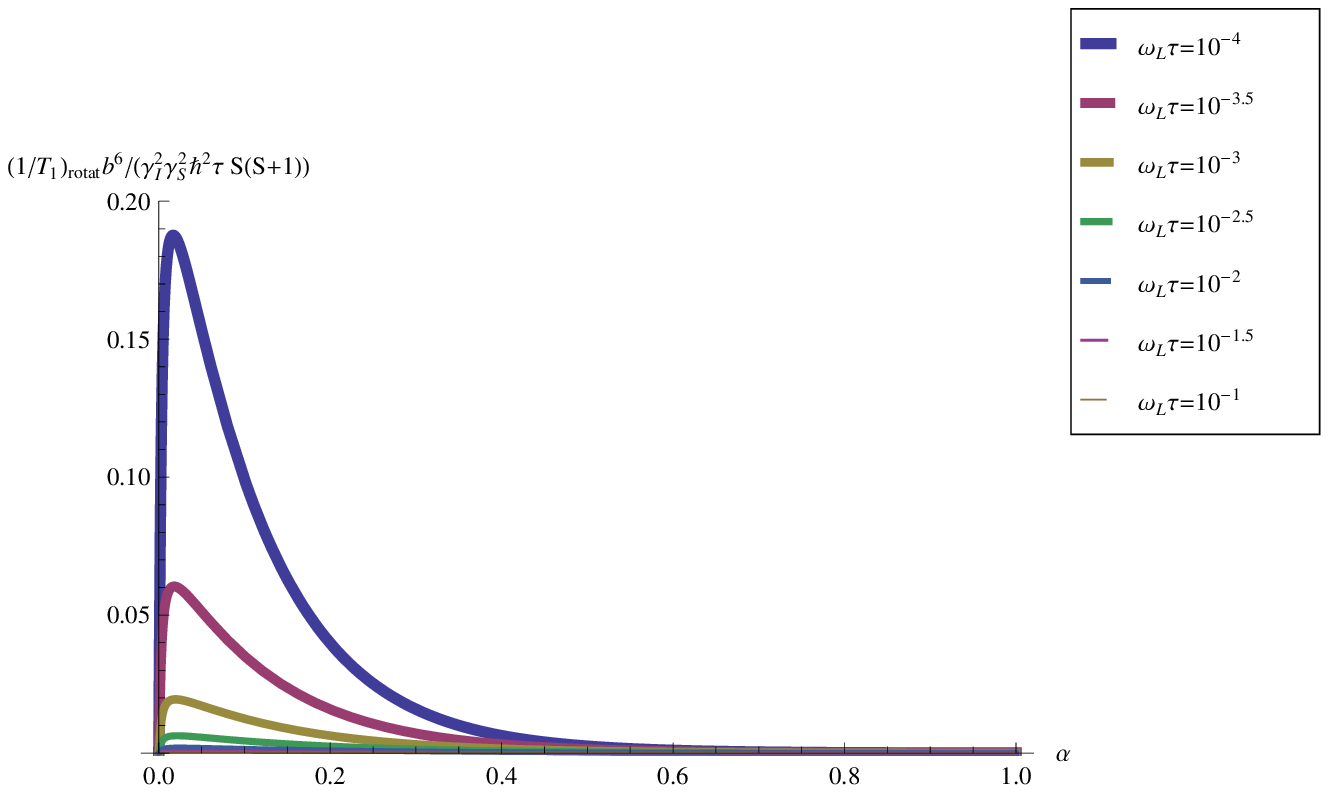}
\end{center}
\caption{Spin-lattice relaxation rate for $^{15}N - H$ nuclear pair of non-identical spins from fractional rotational diffusion for the case of cone axes directed along the magnetic field (eq. (\ref{eq29})) in the cone with $\theta_0=5^\circ$ as the function of the fractional index $\alpha$
at different values of Larmor frequency $\omega_L$: $\omega_L\tau=10^{-4}$ (thick line); $\omega_L\tau=10^{-3.5}$; $\omega_L\tau=10^{-3}$; $\omega_L\tau=10^{-2.5}$; $\omega_L\tau=10^{-2}$; $\omega_L\tau=10^{-1.5}$; $\omega_L\tau=10^{-1}$ (thin line).}
\label{Fig.3}
\end{figure}

\clearpage
\begin{figure}
\begin{center}
\includegraphics* [width=\textwidth] {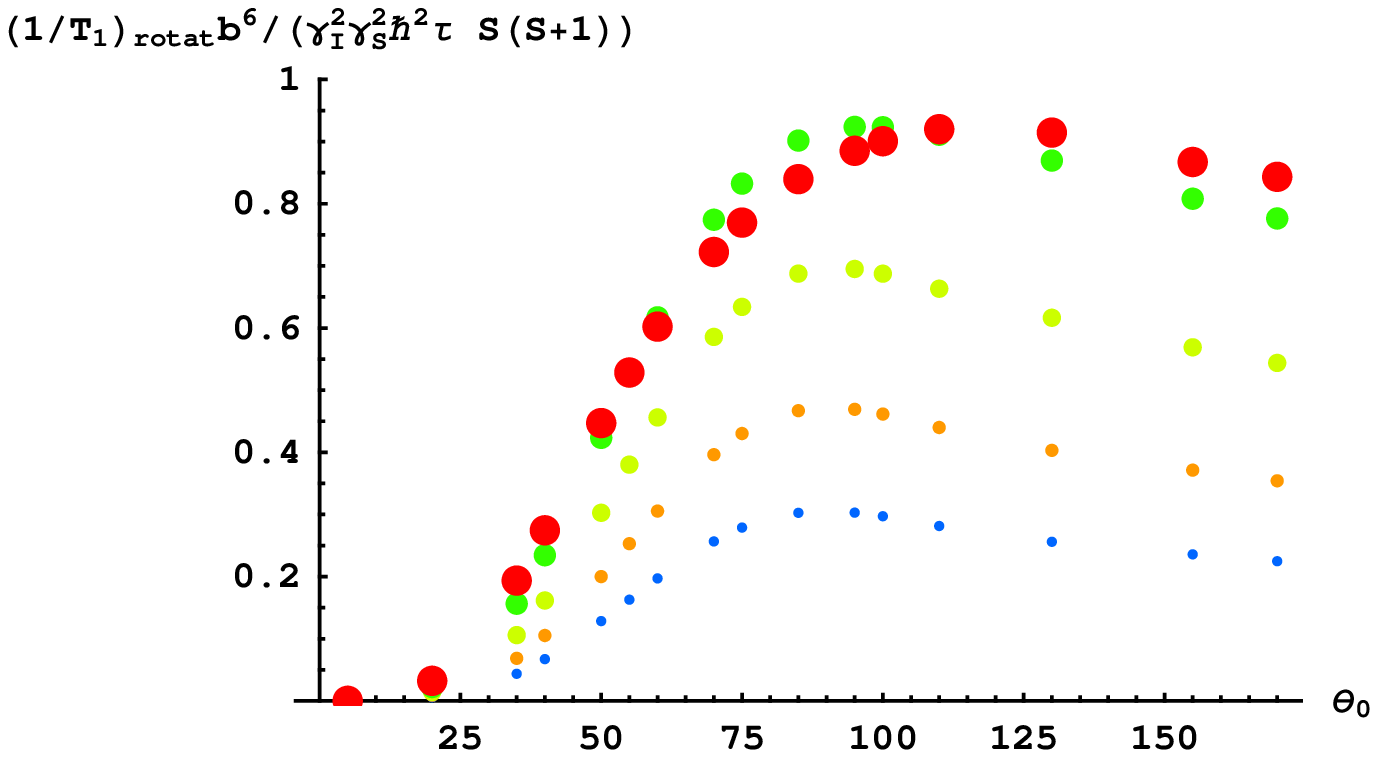}
\end{center}
\caption{Spin-lattice relaxation rate for $^{15}N - H$ nuclear pair of non-identical spins from fractional rotational diffusion for the case of cone axes directed along the magnetic field (eq. (\ref{eq29})) as the function of the cone half-width $\theta_0$ (in degrees) at $\omega_L \tau=0.1$ for
different values of the fractional index $\alpha$: $\alpha =1$ (thin dots); $\alpha =0.8$ ; $\alpha =0.6$ ; $\alpha =0.4$ ; $\alpha =0.2$ (thick dots).}
\label{Fig.4}
\end{figure}

\clearpage
\begin{figure}
\begin{center}
\includegraphics* [width=\textwidth] {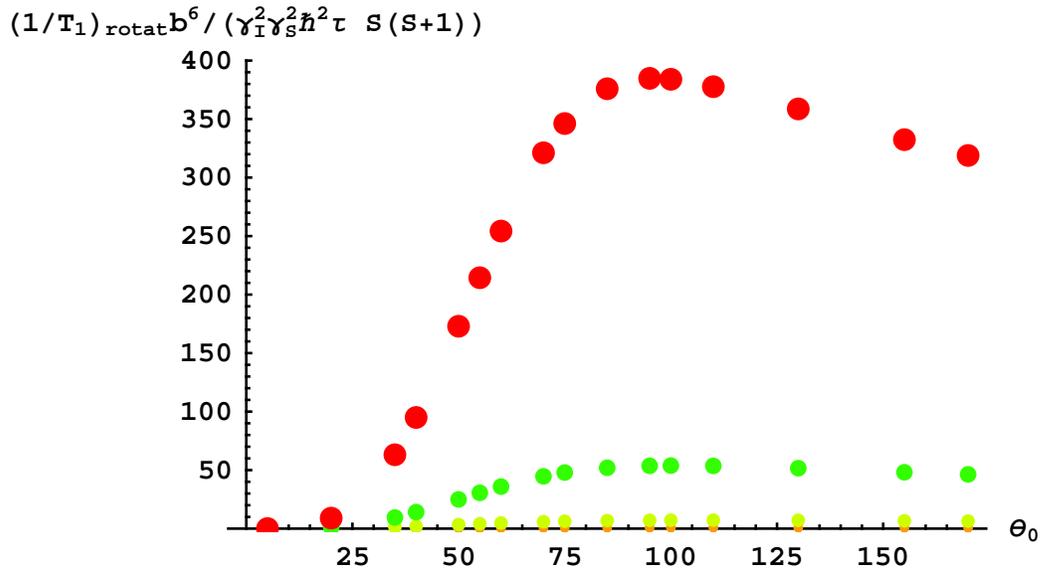}
\end{center}
\caption{Spin-lattice relaxation rate for $^{15}N - H$ nuclear pair of non-identical spins from fractional rotational diffusion for the case of cone axes directed along the magnetic field (eq. (\ref{eq29})) as the function of the cone half-width $\theta_0$ (in degrees) at $\alpha=0.2$ for
different values of the Larmor frequency: $\omega_L \tau=1$ (thin dots); $\omega_L \tau=10^{-1}$; $\omega_L \tau=10^{-2}$ ; $\omega_L \tau=10^{-3}$ ; $\omega_L \tau=10^{-4}$ (thick dots).}
\label{Fig.5}
\end{figure}

\clearpage
\begin{figure}
\begin{center}
\includegraphics* [width=\textwidth] {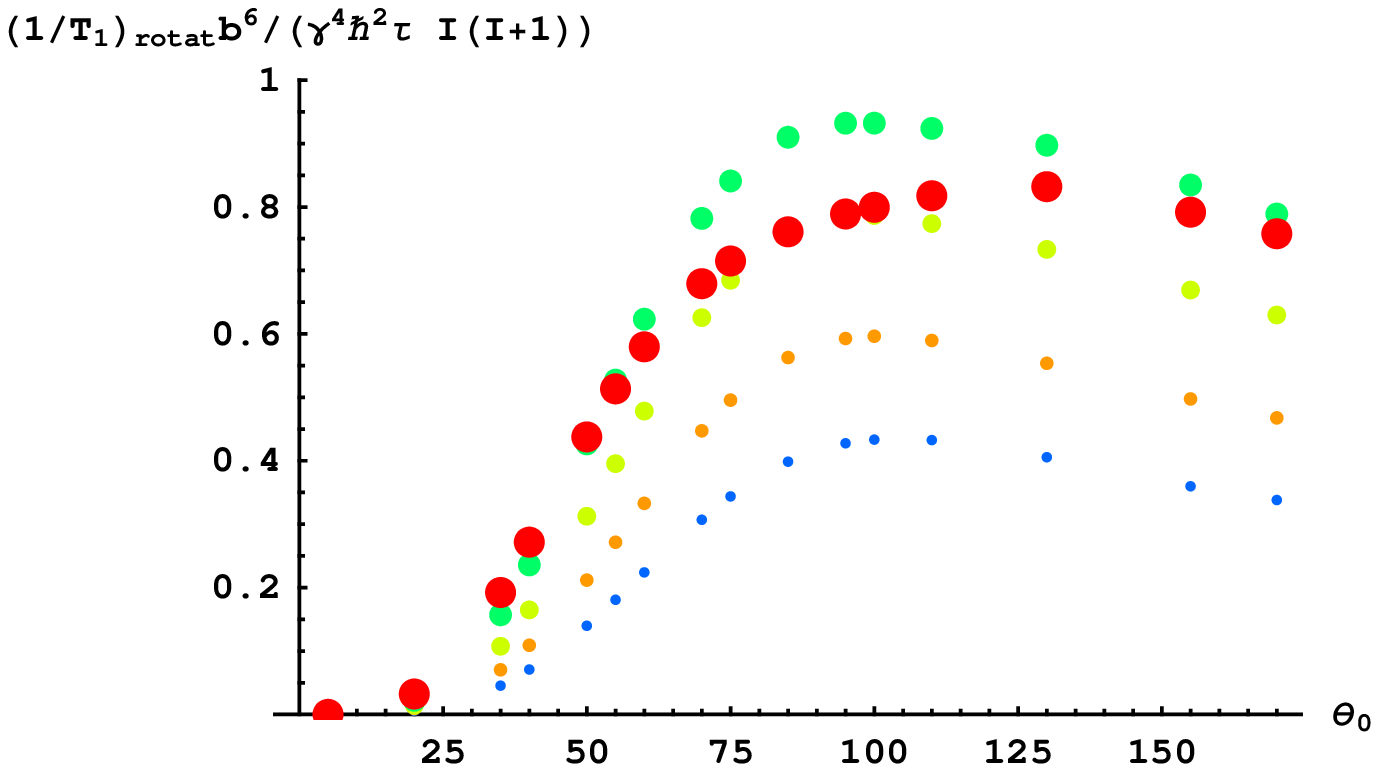}
\end{center}
\caption{Spin-lattice relaxation rate for nuclear pair of identical spins from fractional rotational diffusion for the case of cone axes directed along the magnetic field (eq. (\ref{eq33})) as the function of the cone half-width $\theta_0$ (in degrees) at $\omega_L \tau=0.1$ for
different values of the fractional index $\alpha$: $\alpha =1$ (thin dots); $\alpha =0.8$ ; $\alpha =0.6$ ; $\alpha =0.4$ ; $\alpha =0.2$ (thick dots).}
\label{Fig.6}
\end{figure}

\clearpage
\begin{figure}
\begin{center}
\includegraphics* [width=\textwidth] {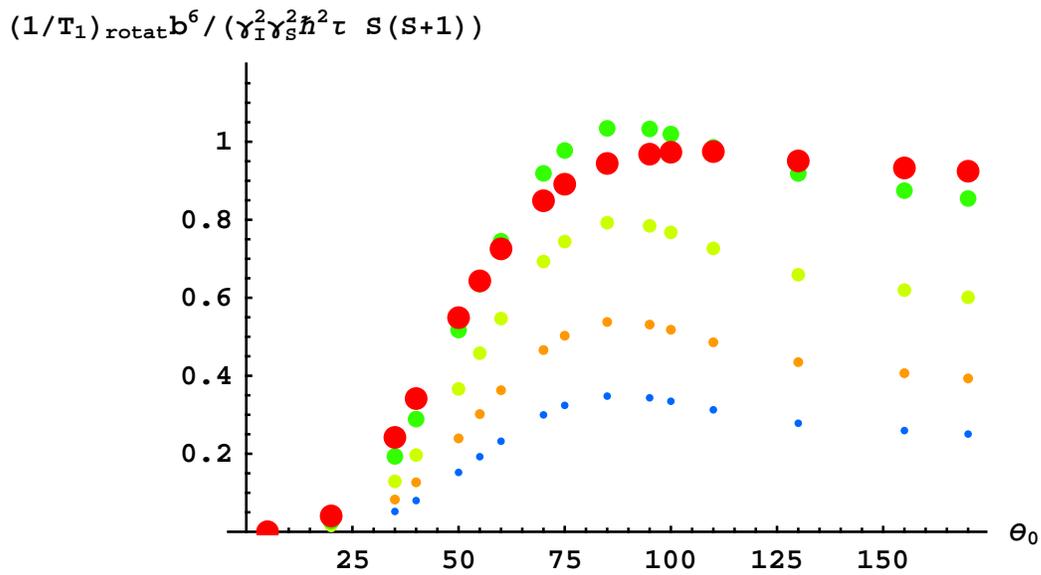}
\end{center}
\caption{Spin-lattice relaxation rate for $^{15}N - H$ nuclear pair of non-identical spins from fractional rotational diffusion for the case of random isotropic distribution (unweighted average $f(\psi)=1$ taking place in powders) of cone axes relative the magnetic field as the function of the cone half-width $\theta_0$ (in degrees) at $\omega_L \tau=0.1$ for
different values of the fractional index $\alpha$: $\alpha =1$ (thin dots); $\alpha =0.8$ ; $\alpha =0.6$ ; $\alpha =0.4$ ; $\alpha =0.2$ (thick dots).}
\label{Fig.7}
\end{figure}

\clearpage
\begin{figure}
\begin{center}
\includegraphics* [width=\textwidth] {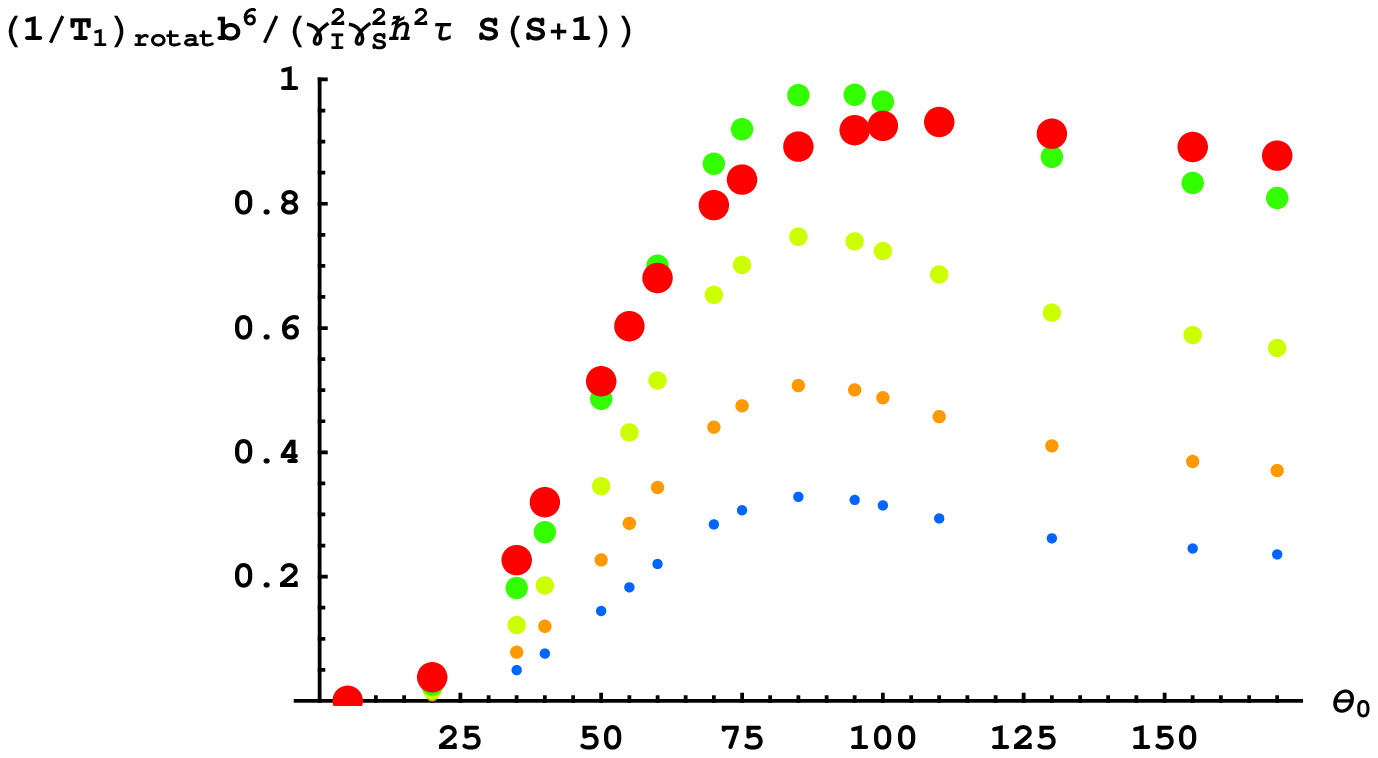}
\end{center}
\caption{Spin-lattice relaxation rate for $^{15}N - H$ nuclear pair of non-identical spins from fractional rotational diffusion for the case of the cone axes to be predominantly oriented along or opposite the magnetic field ($f(\psi)=3\cos^2 \psi$ that may be relevant for, e.g., liquid crystals) as the function of the cone half-width $\theta_0$ (in degrees) at $\omega_L \tau=0.1$ for
different values of the fractional index $\alpha$: $\alpha =1$ (thin dots); $\alpha =0.8$ ; $\alpha =0.6$ ; $\alpha =0.4$ ; $\alpha =0.2$ (thick dots).}
\label{Fig.8}
\end{figure}

\clearpage
\begin{figure}
\begin{center}
\includegraphics* [width=\textwidth] {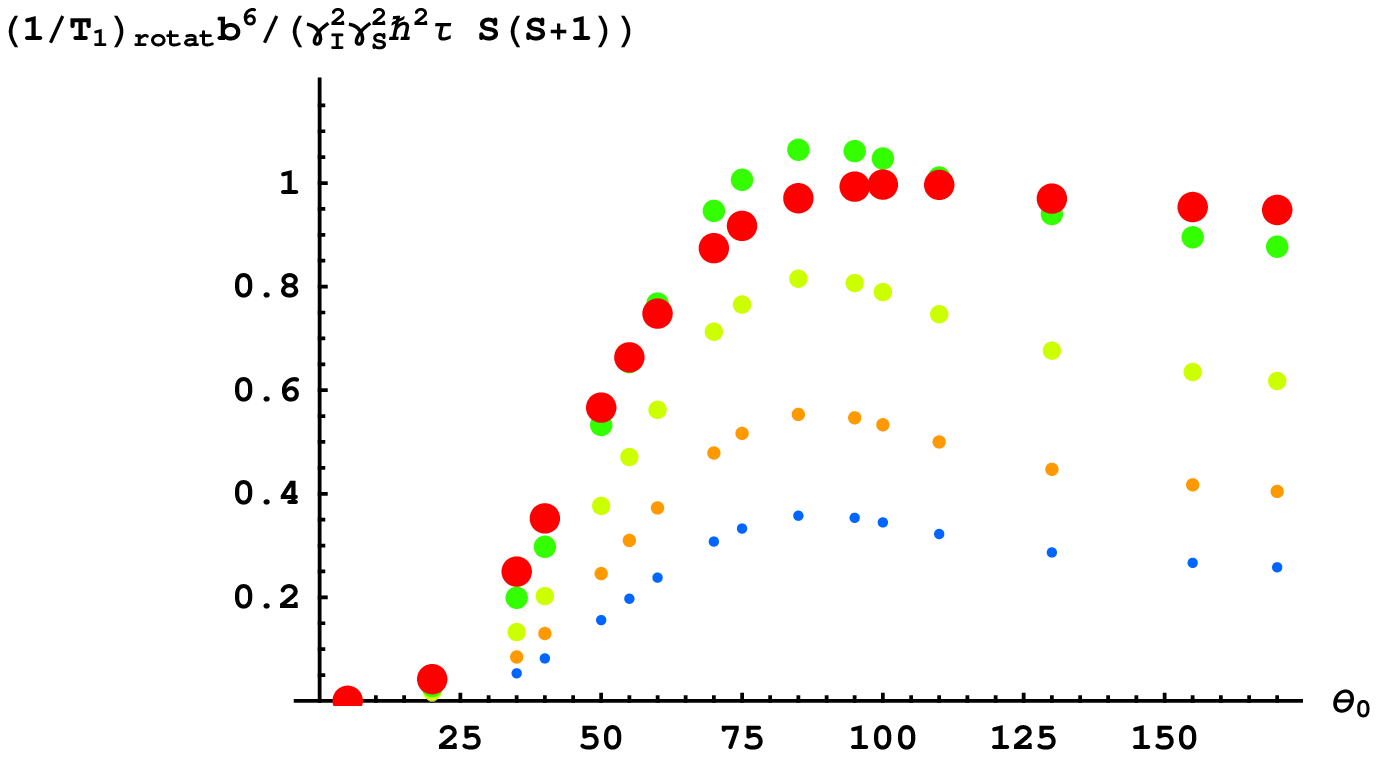}
\end{center}
\caption{Spin-lattice relaxation rate for $^{15}N - H$ nuclear pair of non-identical spins from fractional rotational diffusion for the case of the cone axes to be predominantly oriented transverse to the magnetic field ($f(\psi)=3/2\sin^2 \psi$ that may be relevant for, e.g., liquid crystals) as the function of the cone half-width $\theta_0$ (in degrees) at $\omega_L \tau=0.1$ for
different values of the fractional index $\alpha$: $\alpha =1$ (thin dots); $\alpha =0.8$ ; $\alpha =0.6$ ; $\alpha =0.4$ ; $\alpha =0.2$ (thick dots).}
\label{Fig.9}
\end{figure}

\clearpage
\begin{figure}
\begin{center}
\includegraphics* [width=\textwidth] {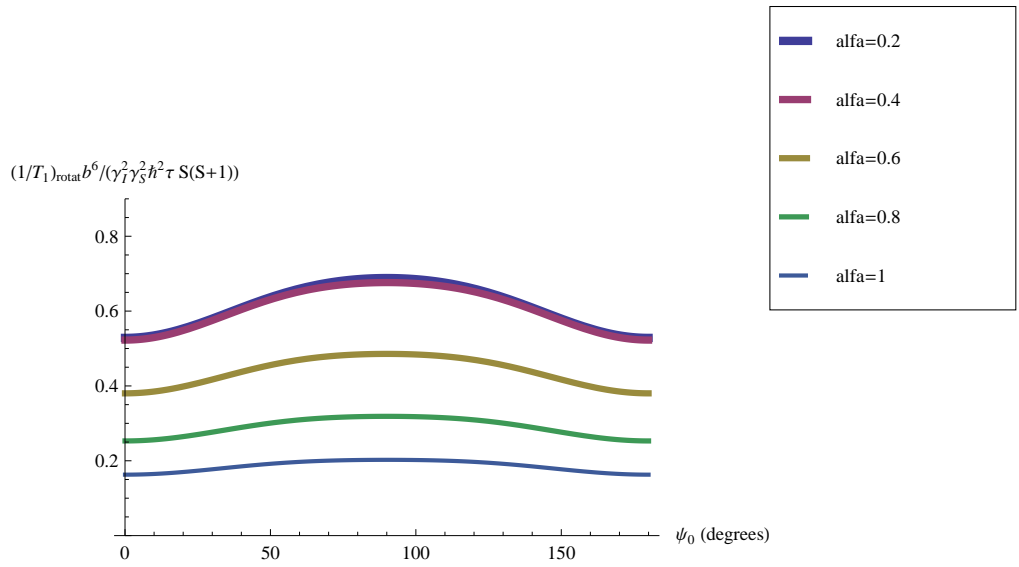}
\end{center}
\caption{Spin-lattice relaxation rate for $^{15}N - H$ nuclear pair of non-identical spins from fractional rotational diffusion in the cone with the half-width $\theta_0=55^\circ$ as the function of the tilt angle $\psi_0$ of the cone axis relative the magnetic field
(corresponding to the distribution function $f(\psi)=\delta (\psi-\psi_0)\left[\cos \left(\psi_0/2\right)\right]^{-1}$)
at $\omega_L \tau=0.1$ for different values of the fractional index $\alpha$: $\alpha =1$ (thin line); $\alpha =0.8$; $\alpha =0.6$; $\alpha =0.4$; $\alpha =0.2$ (thick line).}
\label{Fig.10}
\end{figure}

\clearpage
\begin{figure}
\begin{center}
\includegraphics* [width=\textwidth] {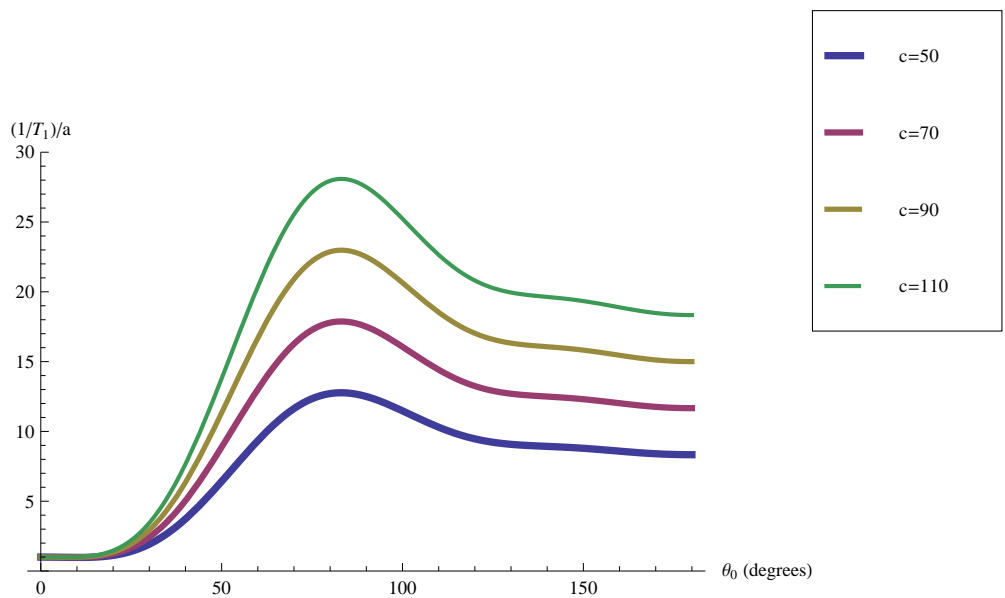}
\end{center}
\caption{Spin-lattice relaxation rate for the model-free approach (eq. (\ref{eq54})) as the function of the cone half-width $\theta_0$ for
different values of the dimensionless parameter $c$ (eq. (\ref{eq53})): $c=50$ (thick line); $c =70$ ; $c =90$ ; $c=110$  (thin line).}
\label{Fig.11}
\end{figure}

\end{document}